\documentclass{mn2e}
\usepackage{graphicx}
\def \sanch{S\'anchez-Bl\'azquez}

\def \kms {${\rm{km}\,\rm{s}^{-1}}$}

\def \z {\phantom{0}}

\def \ha   {H$\alpha$}
\def \hb   {H$\beta$}
\def \hg   {H$\gamma$}
\def \hd   {H$\delta$}

\def \oii   {[O{\sc ii}] 3727}
\def \oiig   {[O{\sc ii}]}
\def \oiii   {[O{\sc iii}]}

\def \oiiib   {[O{\sc iii}] 5007}

\def \nii     {[N{\sc ii}]}
\def \niib    {[N{\sc ii}] 6583}

\def \afe  {[$\alpha$/Fe]}
\def \zh  {[Z/H]}
\def \feh  {[Fe/H]}
\def \nat {_{\rm Nat}}
\def \lick {_{\rm Lick}}
\def \apjl{ApS}
\def \apjs{ApL}

\def \apj{ApJ}
\def \aj{AJ}
\def \mnras{MNRAS}

\def \aanda{A\&A}
\def \z{\phantom{0}}
\title[Shapley Supercluster: Stellar population trends]{
A deep AAOmega survey of low-luminosity galaxies in the Shapley Supercluster: Stellar population trends
}
\author[Russell J. Smith et al. ]
{Russell J. Smith$^{1}$\thanks{E-mail: russell.smith@durham.ac.uk}, John R. Lucey$^{1}$ and Michael J. Hudson$^{2}$\\
$^1$Department of Physics, University of Durham, Durham DH1 3LE, United Kingdom\\
$^2$Department of Physics and Astronomy, University of Waterloo, Waterloo, Ontario N2L 3G1, Canada}
\date{Accepted 2007 July 12.}
\pagerange{\pageref{firstpage}$-$\pageref{lastpage}} \pubyear{2007}
\begin{document}
\label{firstpage}
\maketitle
\begin{abstract}
We present new optical spectroscopy for 342 $R<18$ galaxies in the 
Shapley Supercluster (and 198 supplementary galaxies), obtained from 8 hour integrations with
the AAOmega facility at the Anglo-Australian Telescope. We describe the observations and measurements of 
central velocity dispersion $\sigma$, emission line equivalent widths and 
absorption line indices. 
The distinguishing characteristic of the survey is its coverage of a very wide baseline in velocity dispersion
(90 per cent range $\sigma=40-230$\,\kms), while achieving high signal-to-noise ratio throughout 
(median 60\,per \AA\ at 5000\,\AA). The data quality will enable estimates of Balmer-line ages
to better than 20 per cent precision even for the faintest galaxies in the sample.  
Significant emission at \ha\ was detected in $\sim$30 per cent of the supercluster galaxies, 
including $\sim$20 per cent of red-sequence members. Using line-ratio diagnostics, we find that the
emission is LINER-like at high luminosity, but driven by star-formation in low-luminosity galaxies. 
To characterise the absorption lines, we use the classical Lick indices 
in the spectral range 4000--5200\,\AA. We introduce a new method for applying 
resolution corrections to the line-strength indices.
We define a subset of galaxies with very low emission contamination, based on the \ha\ line,
and fit the index-$\sigma$ relations for this subset. 
The relations show the continuation of the familiar trends for giant galaxies into the 
low luminosity regime, with little change in slope for most indices. 
Comparing the index$-\sigma$ slopes against predictions from single-burst stellar 
population models, we infer the scaling relations of age, total metallicity, \zh, and $\alpha$-element
abundance ratio, \afe.  To reproduce the observed index$-\sigma$ slopes, all three parameters
must increase significantly with increasing velocity dispersion. Specifically, we 
recover: Age$\,\propto\sigma^{0.52\pm0.06\pm0.10}$, Z/H\,$\propto\sigma^{0.34\pm0.04\pm0.07}$, and
$\alpha$/Fe\,$\propto\sigma^{0.23\pm0.04\pm0.06}$ (where the second error reflects systematic effects),
derived over a decade baseline in velocity dispersion, $\sigma=30-300$\,\kms. The equivalent slopes for the subset
of galaxies with $\sigma>100$\,\kms\ are similar for age and Z/H. For $\alpha$/Fe, a steeper slope is recovered
for the high-$\sigma$ subset, $\alpha$/Fe\,$\propto\sigma^{0.36\pm0.07}$. 
The recovered age$-\sigma$ relation is shown to be consistent with the observed evolution in the giant-to-dwarf
galaxy ratio in clusters at redshifts $z=0.4-0.8$. 
In a companion paper, we determine the age, \zh\ and \afe\ 
for individual galaxies, and investigate in detail the distribution of galaxy properties at fixed 
velocity dispersion. 
\end{abstract}

\section{Introduction}

A key challenge in the study of galaxy formation is to explain the emergence and build-up
over cosmic history of the tight red sequence of non-star-forming galaxies, especially in 
the richest environments. At the present epoch, red sequence galaxies account for
some 50 per cent of the total stellar mass (e.g. Bell et al. 2004). 

The evolution of the red galaxy luminosity function provides a direct probe of the mass-assembly history 
of the red sequence. Recent surveys suggest growth in the red sequence mass by factors of 2--3 since redshift $0.7-1.0$, 
dominated by galaxies with relatively low luminosity ($\la{}L^\star$), while the most luminous red galaxies show little number evolution 
over the same interval (e.g. Bell et al. 2004; Faber et al. 2005; Bundy et al. 2006; Wake et al. 2006; Brown et al. 2007; Scarlata et al. 2007).
The history of star formation is in general distinct from the mass-assembly history of the galaxies, since
dissipationless (`dry') mergers of gas-poor systems merely redistribute existing stars (e.g. van Dokkum 2005). 
Faint field-galaxy surveys reaching to high redshifts reveal that star formation has shifted from high-mass galaxies in 
the past to being preferentially active in low-mass systems today, a trend referred to as `downsizing' 
(e.g. Cowie et al. 1996; Juneau et al. 2005). 

In galaxy clusters, the Butcher--Oemler effect (Butcher \& Oemler 1984) constituted 
some of the earliest evidence for the recent rapid growth in the red sequence: a population of blue star-forming galaxies 
existed in some rich clusters at $z\sim0.3$ which is apparently absent at the present day, having presumably
faded into the quiescent population in the intervening $\sim$3\,Gyr. Rapid evolution in luminosity-selected studies
is driven largely by a decline in low-mass star-forming galaxies, which fade onto the faint end of the 
red sequence (e.g. Holden et al. 2006). After star-formation is `quenched', especially if the remaining gas was consumed in a rapid 
starburst, the transformed galaxies may pass through a K+A phase (e.g. Dressler \& Gunn 1983; Couch \& Sharples 1987). 
The characteristic mass scale of K+A galaxies is observed to be larger at high redshift (e.g. Tran et al. 2003) than in local clusters 
(e.g. Poggianti et al. 2004), a manifestation of downsizing in the formation of the red sequence itself. Many recent studies 
suggest that distant clusters are deficient in faint red galaxies, relative to local analogues (e.g. Kodama et al. 2004; 
de Lucia et al. 2004, 2007; Muzzin et al. 2007; Stott et al. 2007), as expected if many of today's faint red sequence galaxies were indeed 
star-forming at look-back times above a few Gyr (but see Andreon 2006 for contrary evidence). 
If the present day red population is compared naively to the galaxies already on the red sequence at high redshift, 
we inevitably find only modest evolution the stellar populations (the so-called 
`progenitor bias'), since we are not comparing to the full progenitor set (e.g. Kaviraj et al. 2006). None the less, 
recent studies of the red-sequence scaling relations, such as the Fundamental Plane, indicate a faster evolution in the 
stellar populations of lower luminosity galaxies, again consistent with the downsizing scenario (e.g. J\o{}rgensen et al. 2006). 

An alternative means of probing the build-up of the quiescent population is to measure explicitly the stellar population
ages of galaxies on the red-sequence at the present day. A mass dependence in the quenching times should show up in the 
current ages of galaxies, so long as the latter are not {\it in general} contaminated by ongoing or renewed star-formation 
significantly after the primary quenching event. (This assumption might be better justified in cluster environments than in 
the low-density field where rejuvenation through mergers could be more common.) 
Determining the average stellar ages of galaxies from  their integrated light is fraught with
difficulties, from the classic age--metallicity degeneracy (e.g. Worthey 1994) to the presence of non-solar abundance
ratios (e.g. Worthey, Faber \& Gonz\'alez 1992) and the confounding effects of composite stellar populations 
(e.g. Serra \& Trager 2007). Despite these challenges, modern spectral synthesis models, notably those of Worthey (1994) 
and Thomas and collaborators (Thomas, Maraston \& Bender 2003; Thomas, Maraston \& Korn 2004), have enabled great progress
in separating the effects of age, metallicity and abundance ratios on the principal spectral lines. 
(While we will focus here on line-index methods, we note that Heavens et al. (2004) have developed an alternative, 
`full-spectrum', method to constrain star-formation histories from the present-day galaxy population.)
 
Many studies have employed these and similar models to analyse modest samples of low-redshift galaxies, mostly
corresponding to giant elliptical and S0 galaxies. We highlight here only a selection of notable work.
Kuntschner (2000) concluded that Fornax cluster ellipticals form a metallicity sequence at constant (old) age, 
while S0s in Fornax show a wider age spread. Trager et al. (2000b) recovered a wide spread in ages, with an 
age--metallicity anti-correlation maintaining the tight broad-band colour--magnitude relation. They show a tendency
for younger ages at lower $\sigma$, at least for galaxies outside of rich clusters. More recently 
Thomas et al. (2005) compiled data from a number of sources for $\sim$200 galaxies, arguing for a fairly 
shallow age--mass relation (although this conclusion was reached after excluding some young low-mass objects from their fits). 

Complementary to these high signal-to-noise (S/N) studies of small galaxy samples, a number of works have used spectroscopic surveys
of thousands of galaxies to extract the age-mass relations for red galaxies, although with lower typical S/N per galaxy. 
Among these, Nelan et al. (2005), working from the red-sequence-selected NOAO Fundamental Plane Survey (NFPS), strongly confirmed 
the earlier hints at an age--mass trend, with Age$\,\propto\sigma^{0.6\pm0.1}$
(see also Smith et al. 2006). Subsequently, similar trends were obtained by 
selecting red and/or early-type galaxy samples from the SDSS: 
Bernardi et al. (2006) recovered Age$\,\propto\sigma^{0.8-1.2}$ (depending on which age estimator is used), while
Clemens et al. (2006) found Age$\,\propto\sigma^{0.8}$ (our estimate from their figure 10). 
Although these surveys suggest an emerging consensus on the existence of the age--mass relation, there are still some
apparently contradictory results. For instance, S\'anchez-Bl\'azquez et al. (2006b) report uniformly old ages in their
sample of $\sim$20 Coma ellipticals (including five galaxies with $\sigma=30-100$\,\kms), although their Virgo and `field' samples
show a trend with Age$\,\propto\sigma^{0.5-1.0}$. 
It is possible that different selection methods, in particular colour selection versus
detailed visual morphological inspection, are at the root of these differences. 

Given the strong intermediate-redshift evidence for downsizing and the mass-dependent build-up of the red sequence, there
is a pressing need to determine present-day galaxy ages over a wide mass baseline. Among the studies described
above, however, most are limited to galaxies near the break in the luminosity function, $M^\star$. For example, the median
velocity dispersion, $\sigma$, is $\sim$200\,\kms\ for the SDSS-based studies (Bernardi et al., Clemens et al.), and 
for the Thomas et al. compilation (which includes only four galaxies with $\sigma<100$\,\kms). 
The NFPS is somewhat better in this regard, with median $\sigma$ around
150\,\kms, but the S/N achieved on individual galaxies is sufficient only to determine an average age 
at low mass. To trace the build-up of the red sequence since $z\sim1$, where the activity is mostly in faint 
galaxies, it is necessary to extend the sampling to still lower $\sigma$, while maintaining high spectral S/N (required
for reliable age measurements), and samples of meaningful size and reproducible selection criteria. 
A first step towards this goal was made by Mobasher et al. (2001) in their survey of Coma cluster members, extending to low
luminosities. However, the resolution of this survey was not sufficient to measure velocity dispersion to act as a mass proxy, 
and the S/N (per Angstrom) was too low to determine meaningful ages for individual galaxies in the fainter part of the sample. 
Using the Mobasher et al. data, Poggianti et al. (2001) found only a very weak dependence of age on luminosity, and a very wide
spread in age at all luminosities. 
Caldwell, Rose \& Concannon (2003) made a higher-resolution and higher-S/N study of 175 galaxies (many in Virgo), with 40 galaxies
in the regime $\sigma=40-100$\,\kms, recovering a trend of Age$\,\propto\sigma^{0.8}$. 
More recently, Matkovi\'c \& Guzm\'an (2005) have observed $\sim$90 galaxies in Coma, including 37 galaxies with $\sigma<100$\,\kms\
and $S/N>20$\,per Angstrom. This sample appears to follow an age--mass relation with a similar slope to the Nelan et al. result 
(Matkovi\'c, private communication). Finally, two studies of true dwarf galaxies illustrate the wide range in properties
which perhaps co-exist in the low-luminosity population: 
Geha, Guhathakurta \& van der Marel (2003) observed 17 
Virgo dwarf ellipticals (dEs) with $\sigma=23-45$\,\kms, and obtained an average age of $\sim$5\,Gyr. At the other
end of the surface-brightness scale, a sample of six ultra-compact dwarfs in Virgo, also with 
$\sigma\sim30$\,\kms, yields ages of $\ga10$\,Gyr (Evstigneeva et al. 2007).

This paper presents the first results from deep observations of galaxies in the Shapley Supercluster
with the Anglo-Australian Telescope (AAT). The objective is to investigate the not only the mean scaling relationships 
but also the intrinsic scatter in the stellar population parameters, their correlation structure, and their dependence 
on luminosity, morphology and environment. Particular emphasis will be placed on the faint red-sequence galaxies, which are expected to have experienced the most recent truncation events. Such an analysis requires age and metallicity measurements
for hundreds of individual galaxies, spanning a wide range in mass, and reaching the level of precision so far achieved 
only for small samples of high-mass objects. 
The Shapley Supercluster (Proust et al. 2006 and references therein), is an ideal target for such a campaign, since 
it provides an exceptional density of galaxies and clusters within a few square degrees on the sky. The angular scale of the 
supercluster and the density of feasible targets are well matched to the two degree diameter field of the AAOmega fibre-fed 
spectrograph (Sharp et al. 2006). For a sample with these characteristics, AAOmega delivers a larger multiplex advantage 
than any instrumentation available on larger telescopes.

The structure of the paper is as follows. 
In Section~\ref{sec:observe} we describe 
the sample selection (\ref{sec:photom}), 
the spectroscopic observations and data reduction  (\ref{sec:specobs}), 
and measurements of  redshift and velocity dispersion (\ref{sec:sigmas}), 
emission line equivalent widths (\ref{sec:emissmeas}), 
and absorption line indices (\ref{sec:indices}). 
In Section~\ref{sec:emlines} we present statistics of objects showing significant
emission lines. We next define the sample of low-emission supercluster members
for further analysis, and tabulate data for this sample (Section~\ref{sec:finalsamp}).
Section~\ref{sec:ixsigs} presents an initial analysis of the stellar population scaling relations, 
based on the index$-\sigma$ relations. A comparison of these results to previous work is made in Section~\ref{sec:otherwork}, 
and the the main conclusions reviewed in Section~\ref{sec:concs}. 

In a companion paper (Smith, Lucey \& Hudson 2007, hereafter Paper II), we use our line-strength data 
to infer the age, total metallicity and \afe\ ratio for each galaxy, and investigate the age--metallicity--mass relations
and their intrinsic scatter. 

Throughout this paper, we adopt cosmological parameters $(\Omega_M,\Omega_\Lambda,h)=(0.3,0.7,0.7)$.
For reference, at the redshift of Shapley ($z=0.048$), one arcminute corresponds to 57\,kpc, 
and the distance modulus is $m-M=36.65$. 

\section{Observations and parameter measurements}\label{sec:observe}

\subsection{Sample selection}\label{sec:photom}

The primary galaxy sample was selected using $BR$ imaging data from the NOAO
Fundamental Plane Survey (NFPS), described by Smith et al. (2004). The 
images were obtained with Mosaic-II at the Cerro Tololo Inter-American Observatory 4m 
telescope in March 2000. Targets were selected from the same photometric catalogues,
generated using SExtractor (Bertin \& Arnouts 1996), as were used to defining 
the NFPS spectroscopic sample. The systematic zero-point errors in the catalogues
are estimated at $\sim$0.03\,mag. Star-galaxy separation is via the usual stellarity 
index of SExtractor, limited by image quality ranging from 0.9 to 1.5\,arcsec FWHM. 

\begin{figure}
\includegraphics[angle=270,width=85mm]{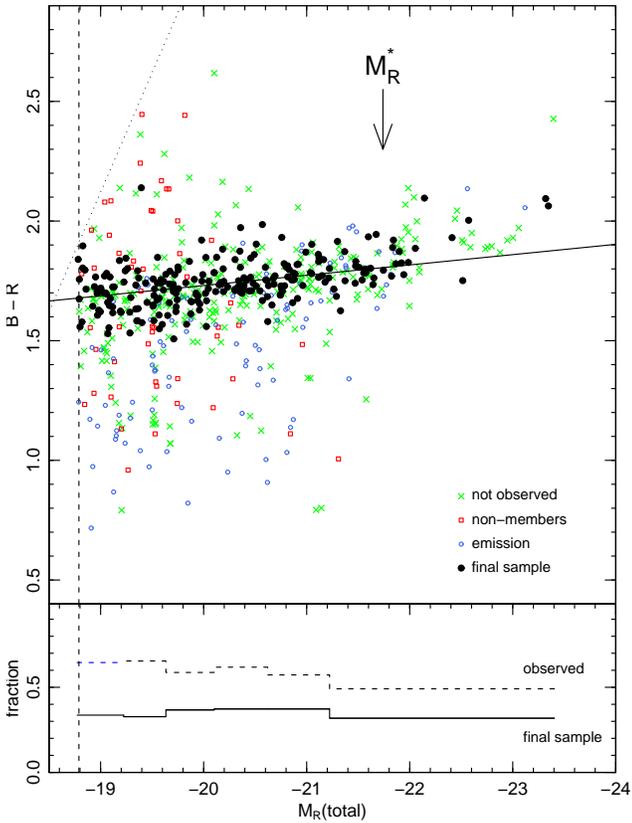}
\caption{Sample selection from the optical colour magnitude relation using NFPS photometry.
Colours here are differences of total magnitudes, not true matched-aperture colours. 
The data are shown after correction for galactic extinction. 
The filled circles show the galaxies which are used in the absorption-line analysis (see Section~\ref{sec:finalsamp}); 
open circles show galaxies in the redshift range adopted for Shapley, but rejected from our sample based on emission at \ha. 
Open squares represent foreground and background galaxies based on our spectra, while crosses denote objects in the NFPS 
imaging regions but to which spectrograph fibres were not assigned.
The dashed and dotted lines show the photometric catalogue limits of $R<18$ and $B<20$, 
for the median galactic extinction in the sample, while the arrow shows the characteristic luminosity $M^\star$
for red galaxies in the SDSS (Baldry et al. 2004).
The solid line shows a robust fit for the emission-free supercluster members. 
In the lower panel, we show the 
fraction of $R<18$ galaxies observed, relative to the NFPS photometric catalogue (dashed line), and the fraction
of NFPS catalogue galaxies which are used in the absorption-line analysis (solid line).
}\label{fig:brcmr}
\end{figure}

The input catalogue for the spectroscopic phase was defined purely by total magnitude
with $R<18$ (one magnitude deeper than NFPS), before correction for galactic extinction 
(median $E(B-V)=0.053$, $A_R=0.14$, from Schlegel, Finkbeiner \& Davis 1998). No cuts were made on the colour relative 
to the red sequence (unlike NFPS), nor on morphological type. 
No redshift information was used in selecting the sample, so some contamination by 
background and foreground galaxies is expected. 
The input catalogue is inevitably incomplete with 
respect to low surface brightnesses (SB), since very low-SB galaxies will not have been detected
by SExtractor. Moreover, very compact galaxies will be misidentified as stars in the imaging data available for 
selection (M32, placed at the distance of Shapley, would have an effective radius of 
$\sim$0.1\,arcsec). 
Figure~\ref{fig:brcmr} shows the target catalogue in a $B-R$ versus $M_R$ colour--magnitude diagram. 
Pre-empting the following sections, the figure distinguishes the galaxies to which fibres were ultimately assigned, 
and also those which prove to be within the supercluster and have pure absorption-line spectra.

The NFPS imaging data were obtained in $\sim$40$\times$40\,arcmin$^2$ tiles centred on the three rich clusters
Abell 3556, Abell 3558 and Abell 3562 (for comparison, the virial radius of a $\sigma_{\rm cl}=1000$\,\kms\
cluster is $\sim45$\,arcmin at the Shapley distance). As such, they cover only a small part of AAOmega's two degree 
diameter field of view. Due to fibre-placement limitations, the target galaxies do not exhaust the available fibres. 
Supplementary objects were added from the 2MASS Extended Source Catalogue (Jarrett et al. 2000), which is 
limited approximately by $J<15$, equivalent to $R\la15.6$ for red galaxies. These galaxies are more extended over the field, 
and could readily be assigned unused fibres. However, to maintain a very simple sample description, 
the 2MASS targets will not be analysed in this paper.

\subsection{AAOmega spectroscopy}\label{sec:specobs}

Spectroscopic observations were made with the AAOmega system (Sharp et al. 2006) on the 3.9m 
Anglo-Australian Telescope, on the nights of 2006 April 26 and 29. 
AAOmega consists of the 
2dF fibre positioner (Lewis et al. 2002), linked to an efficient and stable bench-mounted dual-beam spectrograph. 
The positioner can place 392 fibres, with 2\,arcsec projected diameter, within a  $\sim$2\,degree diameter field of view. 
Within each configuration there is a minimum target separation of $\sim$30\,arcsec imposed by the physical size of the 
fibre buttons (for further details of the fibre positioning constraints, see Miszalski et al. 2006).

Two fibre configurations were employed, with substantially overlapping fields of view. The allocation
of fibres to the targets was performed in a two-stage process. First, approximately 
70 fibres known to exhibit significant fringing effects were withdrawn, 
and the remaining `good' fibres allocated to the $R<18$ sample galaxies. Some 30 `good'
fibres were assigned to random positions to sample the background sky spectrum. 
In a second step, the locations of the these fibres were held fixed, while 
remaining fibres (both `good' and `fringed') were allocated  where possible 
to extra targets from the 2MASS sample.

In the blue arm of the spectrograph, the 580V grating was used, delivering a nominal
resolution of 3.5\AA\ FWHM from 3700--5800\,\AA, sampled at 1.0\,\AA\ per pixel. 
The blue data thus cover most of the standard Lick absorption lines, and also the OII 3727\AA\ emission line 
when present. Simultaneous observations were obtained in the red arm of the spectrograph, using the 
1000R grating, in order to measure nebular emission at H$\alpha$.
The instantaneous wavelength coverage is $\sim$1100\,\AA\ and the nominal resolution is 1.9\,\AA, 
sampled at 0.6\,\AA\ per pixel. 
The two fields were observed alternately, in visits of 1.5--3.0 hours, to a total
integration time of 8--9 hours. 
In this way, each configuration was observed always with the same field plate and fibre bundle, 
so that all the spectra for a given galaxy were obtained with the same fibre. (Exceptions 
are seven objects common to both fields, of which one is a star).
Between the two nights, we introduced a small shift ($\sim$30\,\AA) in central wavelength for the blue spectra, in order
to fill gaps left by blocks of bad columns in the detector. In the red arm, a larger shift ($\sim$350\,\AA) was used
to expand the total wavelength coverage (to 5800--7300\,\AA).
A total of 565 targets were observed, with 416 from the primary NFPS sample and 149 from 
the supplementary 2MASS list. 
{Overall, $\sim$60\% of the NFPS sample targets were observed. Note that because the magnitude information
was not used in prioritising fibre assignment, the observed fraction is not a strong function of luminosity.} 

The data were reduced using a combination of the AAO {\sc drcontrol} pipeline with custom-written 
procedures. The two-dimensional frames were first cleaned of manually-identified bad columns, while
remaining bad pixels and cosmic ray hits were identified using {\sc LA Cosmic} (van Dokkum 2001). The well-sampled PSF
of AAOmega ensures that cosmic rays are clearly distinguished from narrow sky lines and emission lines. 
The spectra were flat-fielded and extracted (using a Gaussian profile extraction) with {\sc drcontrol}. 
The combination of spectra from the many 30\,minute exposures is automated, but handled on a galaxy-by-galaxy basis. 
Examination of the data revealed sizable variations in signal-to-noise ratio (S/N) between exposures, for some (but not all) 
objects. 
To lessen the impact of these variations, we first determine the maximum S/N attained for the galaxy, then discard any individual 
exposures with S/N less than a quarter of this maximum value before combining to form the final spectrum. 
{This criterion rejects only $\sim$2.5\% of the $\sim$10000 individual spectra obtained, but $\sim$10\% of galaxies have
at least one rejected spectrum. Around half of the affected spectra are from fibres close to the edge of the field-of-view, while
the other half are concentrated in an annulus around 0.27\,degrees from the field centre. The S/N variations near the 
edge of the field are more frequent at high airmass, and probably result from varying atmospheric refraction over the course
of a field visit, which can cause relative positional offsets of $\sim$1\,arcsec in the worst cases. For comparison the sample galaxies 
have effective radii typically 2--4\,arcsec (F. La Barbera, private communication).}
When combining the spectra we propagate the error spectra from {\sc drcontrol}, including missing data flags, 
to produce a final error spectrum for each galaxy. Finally, an approximate relative flux calibration was imposed
using the observed spectrum of a calibration star.

Each of the combined spectra was examined by eye to identify any objects unsuitable for the
subsequent analysis. These are: seven stars, three objects in which the target is clearly 
contaminated by a second galaxy with very different redshift, or by a star, and 15 galaxies whose combined
spectra are affected by sinusoidal continuum variations, attributed to `fringing'.
These objects are rejected entirely from further discussion.
For the remaining 540 `uncorrupted' galaxies, the visual inspection confirms there is sufficient
signal to measure at least a reliable redshift, and that any remaining blemishes (usually 
from very low-level detector defects and visible only in the faintest galaxies) affect only 
localised wavelength regions. 
Among these galaxies, 402 are from the primary (NFPS) sample and 138 from the 2MASS supplementary sample. 

Total integrations range from five hours to nine hours for most 
galaxies. Around thirty galaxies have integrations from two to five hours, after the S/N cut described above.
Finally, six galaxies, observed in both field configurations, have total integration times of 14--17\,hours. 

\subsection{Redshifts and velocity dispersions}\label{sec:sigmas}

Reliable velocity dispersion measurements depend on comparing the galaxy spectra to templates which match
them as closely as possible. Traditionally, a small number of K giant stars have been observed 
through the same instrumental set-up for use as templates. While the red giant branch indeed contributes 
substantially to the integrated V-band light of red-sequence galaxies, there is a comparable contribution from
main-sequence stars near the turn-off (e.g. Fig. 13 of Maraston 2005). The contribution from hotter stars varies 
systematically with wavelength, and also as a function of the age and metallicity of the target galaxy itself.
As a result, a single template star cannot match well over an extended wavelength region. Errors
associated with this `template mismatch' can cause systematic bias in the recovered velocity dispersions. 
Methods have been developed which minimize template mismatch by fitting optimal combinations
of template stars (e.g. pPXF of Cappellari \& Emsellem 2004). Here however, we adopt an alternative approach, 
using as templates a set of synthetic SSP spectra covering a range in metallicity and age. The key requirement is 
that the spectral resolution of the models exceed the instrumental resolution. In this paper we employ the synthesis 
models by Vazdekis et al. (2007), based on the 2.4\,\AA\ MILES spectral library (S\'anchez--Bl\'azquez et al. 2006c).

Redshifts are required in order to select a common rest-frame wavelength interval for the 
dispersion measurements. We determine absorption-line redshifts using the cross-correlation method (Tonry \& Davis 1981),
as implemented in {\sc iraf} as {\sc fxcor}. The whole observed spectrum is used at this stage. 
A high-order spline fit to the spectrum is used to identify strong emission lines, which are interpolated over 
to reduce their influence in the cross-correlation. For radial velocity 
templates, we employ a reduced set of MILES SSP spectral models. Specifically, we used solar-metallicity
SSP models with ages 0.5, 1.0, 2.0, 4.0, 8.0 and 16.0 Gyr, adopting the redshift corresponding
to the best-matching SSP, as judged from the {\sc fxcor} R-value. The young SSP models proved useful in obtaining
good absorption-line redshifts from galaxies with substantial emission plus strong underlying Balmer 
absorption. We manually inspected all spectra for which the best R-value was less than 10, and those with substantial
variation in the redshifts derived from young and old templates. In all cases, the measured redshifts were confirmed.
Comparing to redshifts from the NFPS (Smith et al. 2004), for 124 galaxies in common, we obtain a median offset of 
11\,\kms\ (the new measurements larger), with a scatter of 23\,\kms. 

As a preliminary to measuring velocity dispersions, we determine the delivered resolution of our spectra as a function 
of wavelength, by comparing the spectra obtained for a K2 giant, HR7149, with the spectrum for this star in the
Indo-US stellar library (Valdes et al. 2004). For wavelengths of 4500--5750\,\AA, we find a constant 
3.0\,\AA\ FWHM smoothing must be applied to the Indo-US spectrum for optimal match to the AAOmega 
spectrum. Adding in quadrature the Indo-US resolution (described as 1.2\,\AA\ FWHM with 
`small variations'), the AAOmega resolution is thus estimated to be 3.2\,\AA\ FWHM (equivalent to $\sigma_{\rm inst}=82$\,\kms). 
For 4000--4500\,\AA,  the resolution is slightly poorer, $\sim$3.7\,\AA. 
Considering the range where the resolution is approximately constant, and the redshift range spanned by our
principal targets in Shapley, we select the spectral range 4400--5400\,\AA\ in the rest frame, for determination
of the velocity dispersion. This range is fully covered for 461 galaxies with $cz\leq22000$\,\kms.
For velocity dispersion templates, we again employ the MILES SSP models. In the 
wavelength range of interest, the MILES resolution is 2.4\,\AA\ FWHM (figure 4 of S\'anchez--Bl\'azquez et al. 2006c).
We must therefore smooth the MILES template spectra by the quadrature difference (2.1\,\AA), to approximate a
zero-velocity SSP as observed by AAOmega. 

Dispersions were measured using {\sc fxcor} again, calibrating the cross-correlation peak widths to $\sigma$ 
using results from artificially broadened templates (e.g. see Wegner et al. 1999). 
The trial SSPs form a grid with metallicities \feh = --0.68, --0.38, 0.00, +0.20, and ages 
1.0, 1.4, 2.0, 2.8, 4.0, 5.6, 8.0, 11.2, 16.0 and 17.8 Gyr.
As for the redshifts, we select the result corresponding to the best-matching SSP template. 
Examination of the results for different templates shows that if we had used a single old, solar-metallicity SSP 
template, the dispersions would be strongly biased relative to the `best' SSP, especially for low-mass galaxies.
Specifically, while for $\sigma>100$\,\kms\ the two estimates agree with an offset of only $\sim$2 per cent, and a scatter
of $\sim$11 per cent, for $\sigma=75$\,\kms, the median offset is 6 per cent, and for $\sigma=50$\,\kms, 
the bias is 18 per cent. In these cases the `best' $\sigma$ is on average smaller than the $\sigma$ obtained from a single
old SSP template. 

Uncertainties in the velocity dispersions were estimated using Monte Carlo simulations using the error spectra 
corresponding to each galaxy. From 51 realisations, we numerically sort the resulting measurements and adopt the 
8th and 42nd values as an estimate of the 1$\sigma$ interval. In this process the best-matching SSP model is used 
throughout; errors associated with selecting the template are therefore 
not propagated into the velocity dispersion error.
If the galaxy-versus-model cross-correlation peak width is narrower than the model autocorrelation peak width, 
at the 1$\sigma$ lower limit, we classify the galaxy as kinematically unresolved. This is true for
90 of the total 461 galaxies for which measurements were made. The unresolved galaxies are primarily those with best match to 
very young ($<2$\,Gyr) SSP models; many also have strong nebular emission lines. This probably results largely from
the increased influence of broad A-star Balmer absorption, which degrades our effective instrumental resolution 
for ages younger than $\sim$3\,Gyr.

\begin{figure}
\includegraphics[angle=270,width=85mm]{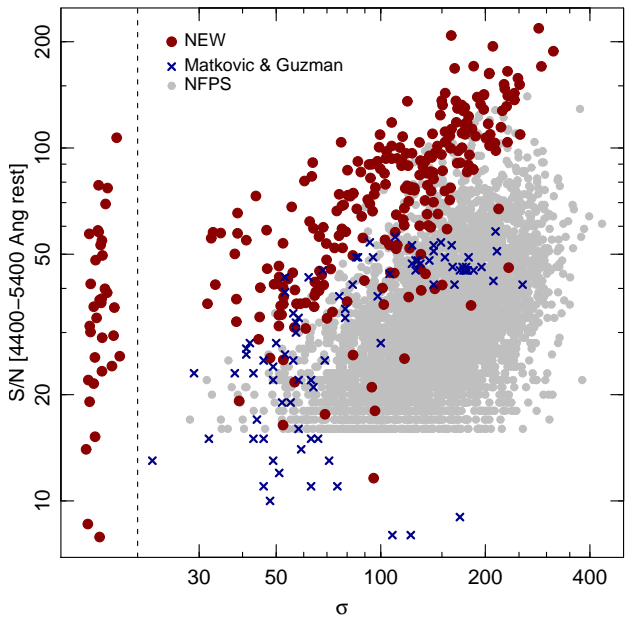}
\caption{Signal-to-noise as a function of velocity dispersion for the dataset presented in this paper 
(filled circles). 
We show only the $\sim$300 galaxies with very low \ha\ emission (see later). 
For comparison, we show the sample of Matkovi\'c \& Guzm\'an (2005) as small crosses,
and the NFPS (Smith et al. 2004) as grey points. The kinematically unresolved galaxies (i.e. 
those with velocity dispersions consistent with zero) are placed at an arbitrary location at the 
left of the panel, separated from the other data by a vertical line. 
}
\label{fig:samplesnr3}
\end{figure}

Figure~\ref{fig:samplesnr3} shows the relationship between measured velocity dispersion and the signal-to-noise 
ratio (S/N, measured over the  4400--5400\,\AA\ rest-frame interval).
The figure suggests that the lowest measurable velocity dispersions in our data 
are $30-50$\,\kms\ (roughly half the instrumental resolution of $\sim$90\,\kms), 
depending on the S/N achieved. 
For comparison, we show the same relation for the entire NFPS spectroscopic dataset
(from Smith et al. 2004). For galaxies in common between the samples, the new data have a factor of $\sim$3.2 
higher S/N ($\sim$10 times deeper). 
A more comparable sample is that compiled for Coma cluster galaxies by Matkovi\'c \& Guzm\'an (2005).
Figure~\ref{fig:samplesnr3} shows that at comparable $\sigma$, the S/N in our spectra is larger by
a factor of around two. 

\begin{figure}
\includegraphics[angle=270,width=82mm]{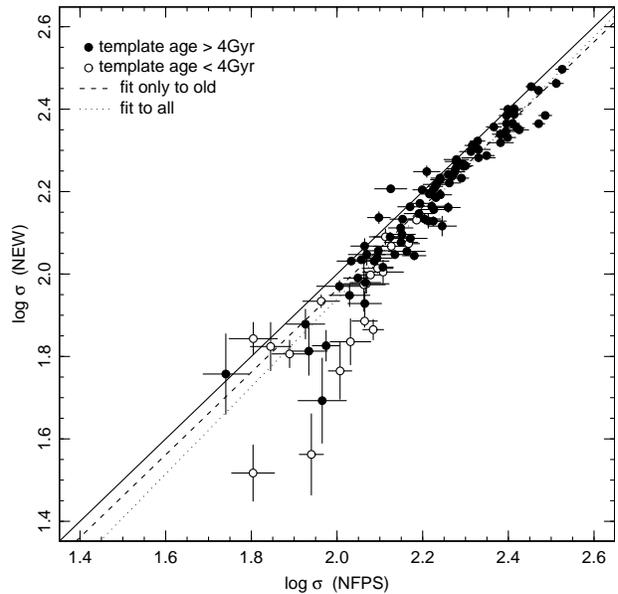}
\caption{Comparison of velocity dispersion measurements with those from NFPS (Smith et al. 2004).
The weighted fit to all points is shown by the dotted line. Filled points
have best-matching templates older than 4\,Gyr; the dashed line is a fit to these galaxies
only. The solid line indicates equality. 
}
\label{fig:logsigvsnfps}
\end{figure}

In Figure~\ref{fig:logsigvsnfps}, we compare the new velocity dispersion measurements with those
from NFPS (Smith et al. 2004), for 105 galaxies in common. No aperture corrections need to be applied, 
since the NFPS fibres are of the same diameter as the AAOmega fibres. The comparison is limited by NFPS to fairly
high luminosity. Moreover there is an explicit bias because low-$\sigma$ galaxies are 
likely to be unresolved in NFPS, and only enter the comparison if errors cause them 
to scatter up to higher $\sigma$. 
We find a median offset of 0.045\,dex ($\sim$11 per cent), with the new measurements 
consistently smaller. The scatter around this offset is 0.068\,dex. For low-dispersion galaxies, 
both the scatter and the average offset increase. The slope of the (error-weighted) fit is 
$1.06\pm0.03$, indicating a significant scale change. 
We attribute this behaviour partly to the improved measurement method which matches the template to each galaxy, 
as compared to using a few K-giant templates as in NFPS. In particular, many of the low-$\sigma$ galaxies are best
represented by young SSP models. As noted above, forcing a fit with a K-giant or old SSP will bias the 
results to higher dispersions in such cases. As a test, we repeat the comparison for only those galaxies
with a best-matching template age greater than 4\,Gyr. The agreement is considerably improved, with a scatter of
0.048\,dex and median offset 0.037\,dex ($\sim9$ per cent). Moreover the scales are consistent (slope 1.00$\pm$0.03).

Hereafter in this paper, we prefer to use the velocity dispersions derived from the best-matching
template. A consequence is that the $\sigma$ scale is slightly stretched relative to the NFPS scale, 
so that measured correlations with $\sigma$ are expected to be flatter than if we had only used K-giant templates.

\subsection{Emission line equivalent widths}\label{sec:emissmeas}

Since our sample has not been selected by morphology or colour, it will contain star-forming galaxies 
as well as passive systems. In addition, it has long been known that 
a substantial fraction even of morphologically-selected ellipticals
exhibit nebular emission lines such as \hb, \oiiib, etc (e.g. Phillips et al. 1986). 
The line ratios may be characteristic of LINER excitation, especially
in more massive galaxies, or of excitation by ongoing star formation, the latter more 
frequently in low-mass galaxies (Nelan et al. 2005; Yan et al. 2005). 

Emission line measurements are necessary to identify passive, non-star-forming galaxies, 
with minimal nebular contamination in the age-sensitive Balmer lines. 
As discussed in Nelan et al. (2005)
and elsewhere, the intrinsic variation among galaxies in the line ratios precludes accurate 
correction for emission infilling based on the \oiiib\ line alone, especially if a wide range in
galaxy mass is being considered. More reliable corrections can, however, be derived if a wider range of 
lines are measured, especially if \ha\ is available (e.g. Caldwell et al. 2003). 

To measure equivalent widths of the principal emission lines, it is necessary first to remove
the underlying stellar continuum, especially in the case of the Balmer lines themselves. 
To this end, we Doppler-corrected each spectrum to the rest frame and divided by the best-fitting 
MILES SSP (as determined from the cross-correlation results), broadened to the measured 
velocity dispersion. Low-order continuum differences were removed using a cubic spline fit. 
Equivalent widths were measured on the ratio spectrum, by direct flux summation over
an interval equivalent to 600\,\kms\ width
{centred at the wavelength expected from the absorption-line redshift. 
This measurement window is wide enough that modest velocity shifts ($\sim$100\,\kms) between gas and stars introduce 
only small variations ($\sim$10 per cent) in the equivalent widths. In Section~\ref{sec:emlines} we will use mainly
the ratios of neighbouring lines, where the effect of such offsets will cancel out.}
Errors in the equivalent widths were estimated from 
the pixel-to-pixel variance in the (template-divided) continuum regions. 

In the blue, the emission lines considered in this paper are \oii, \hb, and \oiiib, while in the 
red spectra, we use \ha\ and \niib. 
{Note that the separation of the latter two lines is equivalent to 900\,\kms, i.e. 
larger than the velocity interval used for measuring the equivalent widths.}
At the redshift of Shapley, \ha\ and \nii\ are affected by atmospheric absorption bands, especially that 
due to O$_2$ at 6870--6955\,\AA. 
Thus we need to disentangle three contributions: stellar continuum, 
nebular emission and atmospheric absorption. To determine the latter, we use the objects with the 
highest S/N, and with large velocity dispersion, for which any high-frequency variations must be 
telluric in origin. The best-fitting SSP models for these galaxies were Doppler corrected to the observed
frame (the opposite of the sense used for continuum removal), broadened to the appropriate velocity
dispersion and divided out of the spectra. The ratio spectra for each galaxy retain the atmospheric
features (always at the same wavelengths) and any emission lines present (at different wavelengths according
to the redshift). Median-combining the ratio spectra effectively removes the emission component, leaving a 
smooth continuum modulated by the telluric absorption. The final step was
to fit out the continuum and create a correction spectrum fixed to unity 
for all pixels below 6800\,\AA. (We have neglected the much weaker absorption features at shorter wavelengths). 
Telluric corrections were derived independently for the two 
fields, since these cover different time intervals. A comparison shows that both the overall amplitude 
and the detailed structure of the O$_2$ band (and the H$_2$O band longwards of 7130\,\AA) 
are very similar between the two fields. 

\subsection{Absorption indices}\label{sec:indices}

The Lick absorption-line indices from HdA\footnote{We use roman type, e.g. HdA for the names
of the conventional line-strength indices, while \hd\ refers to the line itself, whether in absorption or emission.} 
to Fe5406 were measured using {\sc indexf} (see Cenarro et al. 2001), on the flux-calibrated spectra, at the native spectral resolution of $\sim$3.2\,\AA. 
The indices Mg1, Mg2, Fe5270 and Fe5335 are not reported here, since at the redshift of Shapley, these are
contaminated by the 5577\,\AA\ sky emission for a large fraction of the galaxies. 
We use the passband definitions of Trager et al. (1998), and the the conventional units, 
i.e. magnitudes for CN1, CN2 and angstroms for all others. Uncertainties are estimated from the error spectrum
for each galaxy. Before the indices can be compared against spectral synthesis models, three distinct
corrections must be considered: correction to zero-velocity broadening, correction to the spectral resolution of the
models, and correction to match the instrumental response curve used in the models, if this is not a fluxed system 
(e.g. the Lick/IDS system). 

\subsubsection{Velocity broadening corrections}

The measured absorption line indices differ from those computed by spectral
synthesis models, due to broadening by the line-of-sight velocity distribution function. For an 
isolated absorption line, increasing $\sigma$ causes the line to
`leak' from the central index band, depressing the measured index. 
With further broadening, the absorption itself spreads into the 
pseudo-continuum bands, further reducing the measured absorption.
In a complex spectrum with many overlapping features, the effect of velocity broadening
can be assessed only empirically by comparison to artificially-broadened template spectra. 

Most studies in the field have derived velocity broadening corrections (VBCs)
to the indices, based on measurements on artificially broadened spectra of template stars obtained with the 
same instrumentation as the galaxies (e.g. Gonz\'alez 1993; Davies, Sadler \& Peletier 1993; 
Trager et al. 1998). Alternatively, a library of SSP template models can be used to cover a wider range 
in spectral type, and provide a better match to the composite spectra of real galaxies (e.g. Kuntschner 2004). 
The correction described by each template spectrum is usually normalised by the index value at zero velocity dispersion,
and the results from different templates averaged into a single multiplicative correction describing the ratio 
A number of indices (most notably the \hg\ and \hd\ indices of
Worthey \& Ottoviani 1997) can take a range of values which span or 
approach zero. Under such conditions the usual multiplicative correction fails, and 
in these cases an additive correction, based on $I(\sigma)-I(\sigma=0)$ has generally been used.

{
In general, it is not clear whether one should apply an additive or a multiplicative 
correction for any given index.  In the simplest case, where a single correction curve is used for all the target
galaxies, a multiplicative correction may introduce systematic errors as a function of linestrength (and thus indirectly 
as a function of $\sigma$). More recently, many works have derived corrections for each galaxy based on a model
template chosen to match that galaxy (e.g. Kuntschner et al. 2006; \sanch\ et al. 2006a; Kelson et al. 2006). 
In this paper, we adopt an alternative approach to the VBCs, which allows
a general and consistent treatment of all the indices. 
}

Our corrections are based on a range of MILES SSP model templates (Vazdekis et al. 2007), with ages 3--18\,Gyr and
metallicities \feh\ = --0.38, 0.00 and +0.20.
We smooth the template spectra to mimic both the
AAOmega instrumental resolution, and also the degradation caused by velocity broadening in
steps from zero to 500\,\kms. (We neglect any higher order moments of the line-of-sight velocity
distribution, but in principle the method could be generalized to allow for these.) 
The indices are measured on the broadened spectra.
For each step in $\sigma$, the broad coverage of spectral types yields a wide
range in index value. At each value of $\sigma$, we then simply compare the 
index measured on the velocity broadened spectrum, $I\nat(\sigma)$, with that measured only allowing
instrumental resolution, $I\nat(\sigma=0)$. (The subscript `Nat' here refers to measurements at the native
resolution of the spectra.) For most of the standard indices, this comparison yields a 
closely linear relation such that $I\nat(\sigma=0) = a I\nat(\sigma) + b$ (at fixed $\sigma$), with very small 
scatter. This can be applied as a correction for the indices measured on the galaxy 
spectra. The coefficients $a$ and $b$ for each galaxy can be interpolated from the steps in $\sigma$ used
for smoothing the templates. In most cases the SSP models span the range of the measured values. Where this is not the case
(e.g. the highest Mgb5177 values, for $\alpha$-enhanced galaxies), the high degree of linearity in the corrections 
suggests that a small extrapolation is probably safe. In applying the corrections, we account for the effect of the 
linear correction to the index error. The uncertainty in the velocity dispersion (and hence the coefficients $a$ and $b$) 
could also be propagated to the index error, but in practice this contribution is negligible compared to the index error. 

\begin{figure*}
\includegraphics[angle=270,width=180mm]{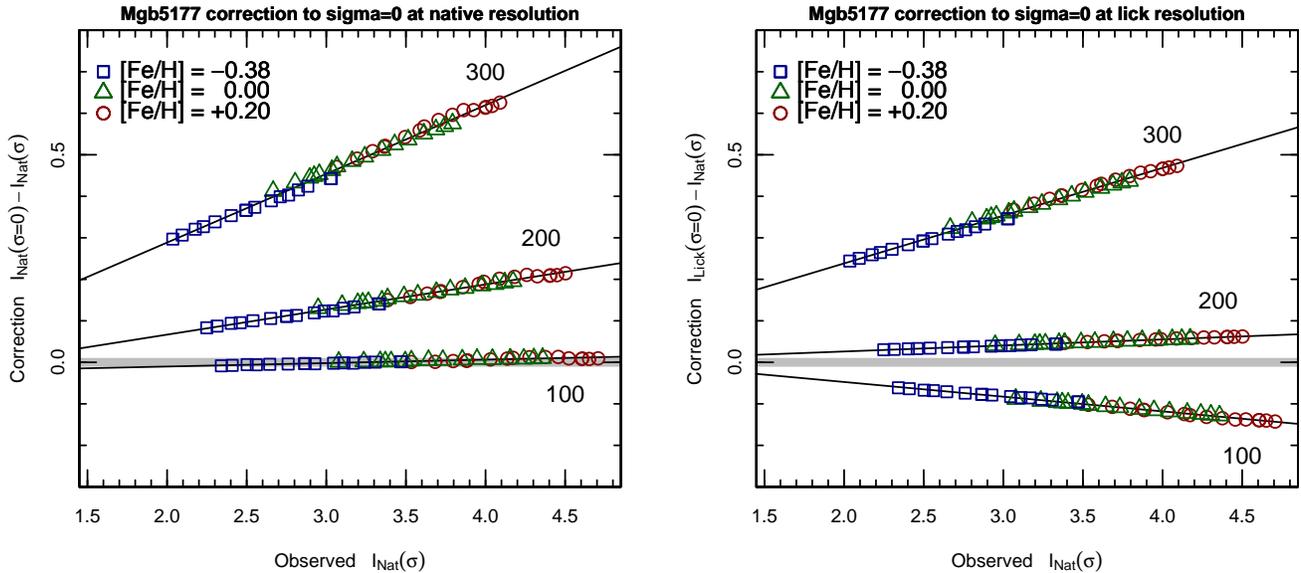}
\caption{Left panel: velocity broadening correction for Mgb5177, correcting to $\sigma=0$ at the native resolution of the spectra.
For each index, we plot $I\nat(\sigma=0) - I\nat(\sigma)$, where $I\nat(\sigma=0)$ is the index measured on the MILES model SSP spectra 
with no velocity broadening (but smoothed to our native instrumental resolution), and $I\nat(\sigma)$ the index measured
after broadening to $\sigma=100, 200, 300$\,\kms, as well as to the native AAOmega resolution.
The latter is thus the index value as `observed'. Points show the corrections for SSPs with \feh=--0.38 (squares), 0.00 (triangles)
and +0.20 (circles). The thick grey line shows zero correction. Solid lines are linear fits to the relation given by the MILES SSPs. 
Right panel: Equivalent corrections to $\sigma=0$ at the Lick resolution. In this
case, $I\nat(\sigma)$ is, as before, the index after broadening by $\sigma$ the native instrumental kernel, while
$I\lick(\sigma=0)$ is measured with the Lick instrumental smoothing, but with no velocity broadening.
}\label{fig:examplevbcs1}
\end{figure*}

\begin{figure*}
\includegraphics[angle=270,width=180mm]{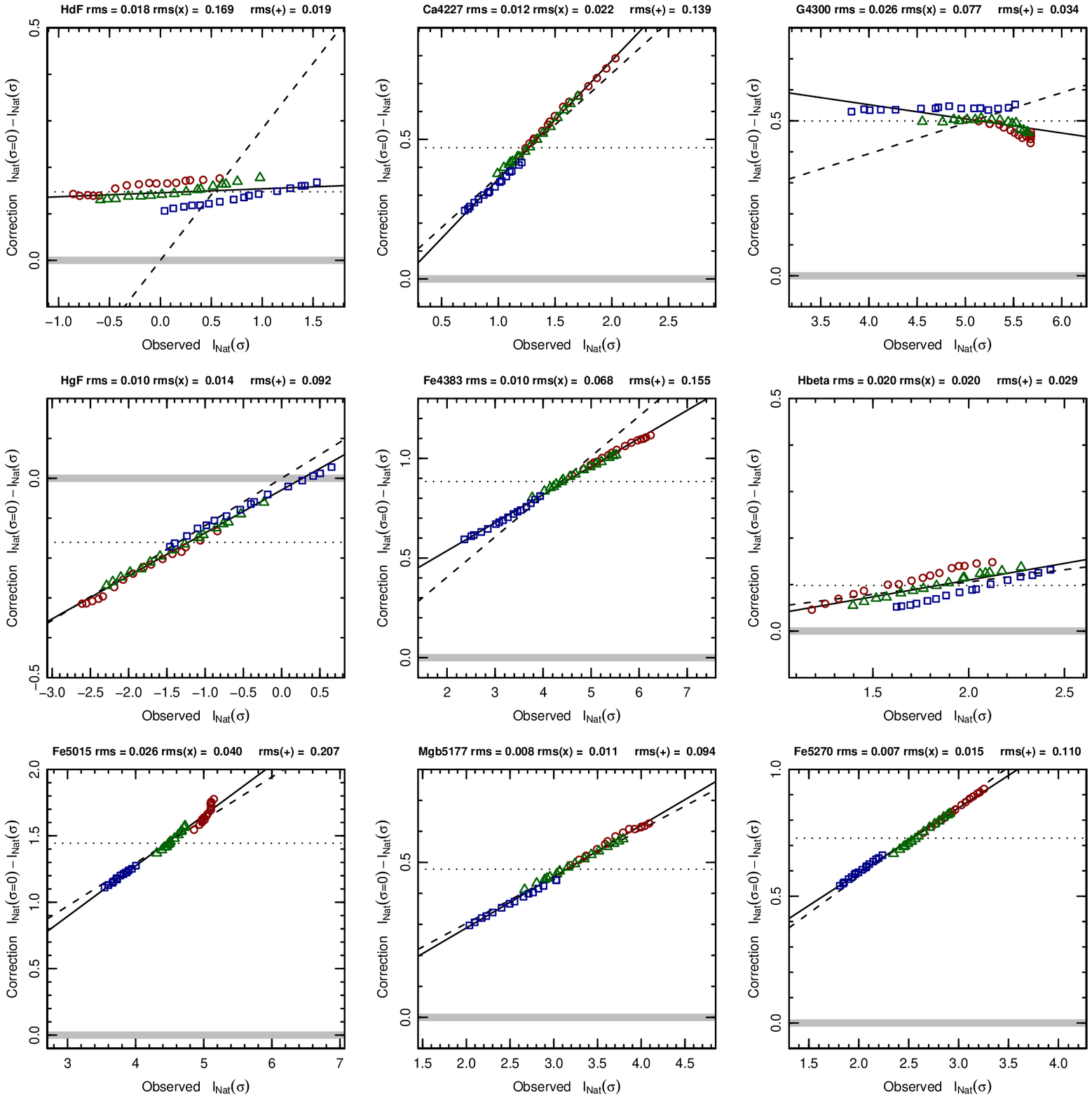}
\caption{Velocity broadening corrections, to the native AAOmega resolution, for a selection of important indices, 
as derived for $\sigma=300$\,\kms. 
The thick grey line shows zero correction. The solid black line is a linear fit to the relation given by the MILES SSPs, i.e. the
correction used in this paper. The dashed line (often very nearly coincident with the solid) shows the best-fitting 
multiplicative correction, i.e. a constant factor  $I(\sigma=0)/I(\sigma=300)$ for all spectral types.
The dotted line shows the best fitting-additive correction, i.e. constant offset $I(\sigma=0)-I(\sigma=300)$ for all spectral
types. 
In the title bar, we note the RMS of the models around the best linear relation, 
as well as the RMS around the multiplicative and additive corrections.
}\label{fig:examplevbcs2}

\end{figure*}

The left panel of Figure~\ref{fig:examplevbcs1} shows the corrections derived for the Mgb5177 index, 
for $\sigma=100, 200, 300$\,\kms. In this case, the correction  $I\nat(\sigma=0) - I\nat(\sigma)$ increases
with index strength such that our correction is almost identical to the usual multiplicative form. 
In Figure~\ref{fig:examplevbcs2} we present the best linear corrections
at $\sigma=300$\,\kms, for a selection of the principal indices, with a comparison to the equivalent 
pure multiplicative and pure additive corrections. The figures demonstrate that the new approach yields linear 
corrections which hold to a high degree of precision over a very wide range of spectral types. 
By contrast, for some indices, the best multiplicative and best additive corrections can be accurate only 
over a limited range of the index value itself.

{ 
As an example, consider the fairly narrow iron index Fe5270. The linear and multiplicative corrections (as usually employed 
for this index) reproduce the observed values over the whole range of SSPs. 
By contrast, in the panel for HdF, we find that the multiplicative correction fails badly, as expected for this index. 
The additive correction, as usually employed, is as successful as our linear fit, although in both cases there is some 
metallicity dependence in the residuals.
In these cases, our results support the form of correction generally adopted. Two counter-examples, where our scheme 
suggests a different form of correction than the standard method, are HgF and G4300. 
For HgF, the traditional additive correction is a poor fit to the broadened spectra. Our linear correction is
much closer to the multiplicative form, but is a better fit than multiplicative for the most Balmer-strong 
(young or metal-poor) models. For G4300, previous work has generally applied a multiplicative correction. 
Our analysis shows that an additive correction is much preferred for this index. However, the index has a complicated
metallicity-dependent behaviour which is not accurately reproduced even by our linear correction scheme. 
}

In summary, the new formulation of the VBCs described here, while not much more complicated or difficult to implement than the 
traditional approach, improves upon the traditional method in at least two regards: (i) greater generality, requiring no 
subjective preference for additive or multiplicative corrections, and (ii) ready applicability over a wide range of 
spectral types for most indices. 
{Our approach should be equivalent to the more modern methods which derive corrections from 
matched template spectra. We contend, however, that describing the VBCs in terms of a mapping from observed index to the value
at $\sigma=0$ is a more natural way to present the correction.}

\subsubsection{Instrumental resolution correction}

In order to compare to models defined at the resolution of the Lick/IDS stellar library,
such as those of Worthey (1994), Thomas et al. (2003, 2004), and Schiavon (2007), it 
is necessary to match the 8--12\,\AA\ wavelength-dependent spectral resolution of the 
IDS instrument. Traditionally this is achieved by artificially broadening the spectra 
prior to measuring the indices. A serious drawback is that this results in redistribution 
of any non-uniform noise in the spectra. For example, in intermediate-redshift observations, residuals from subtracting 
narrow sky lines would be smeared over $\sim$20\,\AA\ intervals. 
In the data described here, noise non-uniformity is less severe, but arises
from the blocks of bad columns in the detector. Since our observed wavelength range was shifted
mid-way through the observations there are (usually) no pixels with unknown values, but in certain regions the 
variance is doubled since only half the integration yields good data. 

An alternative approach, as pointed out by Kelson et al. (2006), is to subsume the resolution correction into 
the same process as the VBCs. The correction to the Lick resolution can be obtained by measuring indices on 
the template SSP models after broadening to the Lick resolution, but without any velocity broadening, denoted $I\lick(\sigma=0)$.
The smoothing required for each index is as tabulated by Schiavon (2007), based on fig. 7 of Worthey \& Ottoviani (1997). 
Corrections from observed to zero velocity dispersion can then be established at the Lick instrumental resolution, 
in the same way as before, by fitting a linear relation $I\lick(\sigma=0) = a I\nat(\sigma) + b$, at fixed $\sigma$
The right panel of Figure~\ref{fig:examplevbcs1} shows the correction for Mgb5177 as an example. Since the Lick resolution
is equivalent to $\sigma\sim200$\,\kms, our higher-resolution measurements are subject to a negative correction at low $\sigma$ and to a positive correction at high $\sigma$. 

To recap, our corrections to zero velocity dispersion and to the Lick resolution are handled in a single step.
The corrections can be applied without smoothing the observed galaxy spectra, correctly 
preserving the noise properties of the data. 

\subsubsection{No calibration to the Lick system}

Finally, the most widely used models of the past few years are those of Thomas et al. (2003, 2004), which are based on
the Lick/IDS stellar library (Worthey et al. 1994). This library is not flux calibrated, instead being defined by the 
wavelength response of an instrument of only historical significance. We did not expend valuable telescope time observing 
Lick standard stars to match this system. Instead, when we compare to the Thomas et al. models we will employ them in a 
relative sense only. This can be done either by considering only the slope of, for instance, the index$-\sigma$ relations
(as in Nelan et al. 2005), or by imposing zero-point shifts to fix a set of the most massive galaxies to match fiducial 
values for age, \zh\ and \afe. Ultimately, we anticipate re-analysing the current data in the context of a new generation of
models now being developed (e.g. Schiavon 2007; Vazdekis et al. 2007) which are tied to carefully 
flux-calibrated stellar libraries (e.g. Jones 1999; Valdes et al. 2004; S\'anchez--Bl\'azquez et al. 2006c). 

\subsubsection{Comparison with NFPS}
\begin{figure*}
\includegraphics[angle=270,width=180mm]{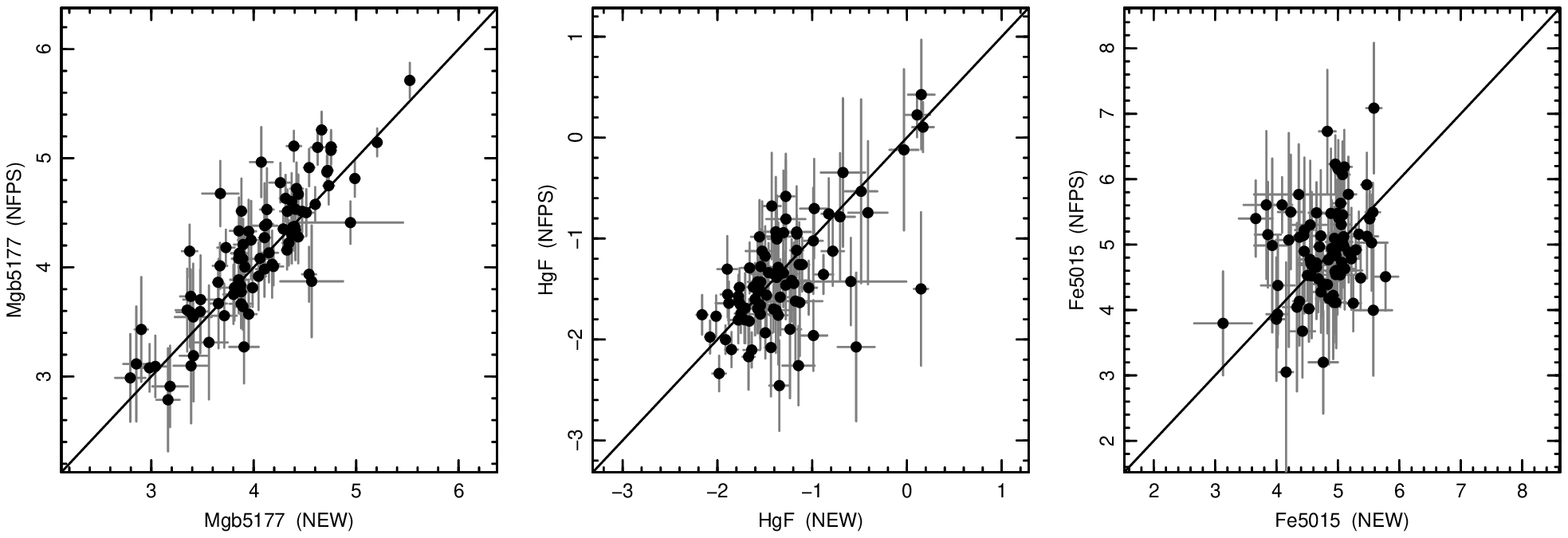}
\caption{Comparison of the new index measurements with data from NFPS (Nelan et al. 2005), for three representative indices.
There are 78 galaxies in common which meet our emission criterion (EW(\ha)$<$0.5\,\AA). The line indicates equality. 
}\label{fig:nfpixcomp}
\end{figure*}
Figure~\ref{fig:nfpixcomp} compares the new index data with that of NFPS (Nelan et al. 2005) for 78 galaxies common to the two
samples. The comparisons are limited by the uncertainties in NFPS, which are typically three times larger than those in the new measurements. 
In the case of Mgb5177, which is well measured in both studies, the data show a zero-point offset of $0.10\pm0.04$.
In the other indices shown, no significant offset is observed. 

\section{Emission line analysis}\label{sec:emlines}

In this section, we analyse the emission line equivalent widths, 
primarily to determine a criterion for selecting an emission-free galaxy sample. 
The incidence of emission lines in cluster members is also 
of interest in itself, reflecting the ongoing consumption of gas in the cluster environment.  

In the following sections, we will use indices measuring the Balmer lines (specifically Hbeta, HgF, HdF) 
as age indicators for the galaxy sample, on the assumption that these are dominated by absorption in stars near 
the main-sequence turn-off. 
If the galaxy contains regions of star formation, or hosts an active galactic nucleus (AGN or LINER), then 
the Balmer lines are contaminated by emission, yielding weaker net absorption and spuriously old age estimates. 
Such galaxies must either be corrected for the effects of emission or excluded from the sample entirely. Many older studies chose 
to correct for \hb\ emission by assuming a constant ratio between the \oiiib\ and \hb\ lines 
(e.g. \hb/\oiii = 0.6, Trager et al. 2000a). However, the measured line ratios, even for red-sequence galaxies, 
span a wide range around this value, and moreover vary systematically with galaxy mass (Nelan et al. 2005; Yan et al. 2006). 
Specifically, correction using \oiii\ under-corrects the emission at low mass and over-corrects at high mass. 
Using \oiii\ alone to select low-emission samples risks the inclusion of high \hb/\oiii\ objects at low $\sigma$. 
Either case could result in a spuriously flattened age$-\sigma$ relation.

A more useful diagnostic, when available, is a measurement of \ha\ emission, since the \ha/\hb\ ratio should be much more stable than, 
for instance \oiii/\hb, which depends strongly on the source of excitation. Moreover, nebular emission at \ha\ dominates over the stellar 
absorption even for very low-levels of star formation or AGN luminosity, providing powerful leverage on the 
contamination of higher-order lines, and robustness against errors in continuum removal (e.g. Caldwell et al. 2003). 

The sample used in this section comprises 342 galaxies selected from the NFPS imaging, with redshifts
compatible with membership of the supercluster (velocities in the range 11670--17233\,\kms, justified in the
following section). \ha\ emission is detected with EW(\ha)$>0.5$\,\AA, for 110 objects i.e. 32\, per cent. 
With this threshold, the detection of \ha\ is always significant at the $>3\sigma$ level. 
Of these 110, emission is also detected with S/N$>$3 in \hb\ for 85 per cent, in \oiiib\ for 89 per cent, in \niib\ for 96 per cent 
and in \oii\ for $\sim$80 per cent. 
The converse proportions are more critical: we are most concerned to ensure
that the measured \hb\ absorption is not substantially contaminated by nebular emission. 
We prefer not to use the directly-measured \hb\ emission to impose a sample cut,
so that errors in selecting the best-fitting continuum template are not propagated directly to the
absorption line measurements. However, we can use these data to assess the reliability of 
selecting according to the other lines. For 232 galaxies without detected \ha\ (`\ha-clean'), only ten (i.e. 4 per cent ) show significant
(S/N$>$3) \hb\ emission. The equivalent fraction for 269 \oiii-clean objects is 10 per cent, 
while for 241 \oiig-clean objects, 12 per cent show \hb\ emission. (In defining clean samples from \oiig\
and \oiii, we did not impose an absolute EW cut comparable to the 0.5\,\AA\ applied for \ha, 
since in these cases there is not a strongly-varying underlying stellar absorption line.)
Given the direct astrophysical connection between \ha\ and \hb, we would have expected
this to be the most {\it accurate} indicator for nebular contamination. The tests presented 
in this paragraph demonstrate in addition that it is also the most {\it sensitive}
indicator from a practical perspective, admitting a factor 2--3 times fewer galaxies with \hb\ emission
than the available lines in the blue spectra. 

\begin{figure*}
\includegraphics[angle=0,width=175mm]{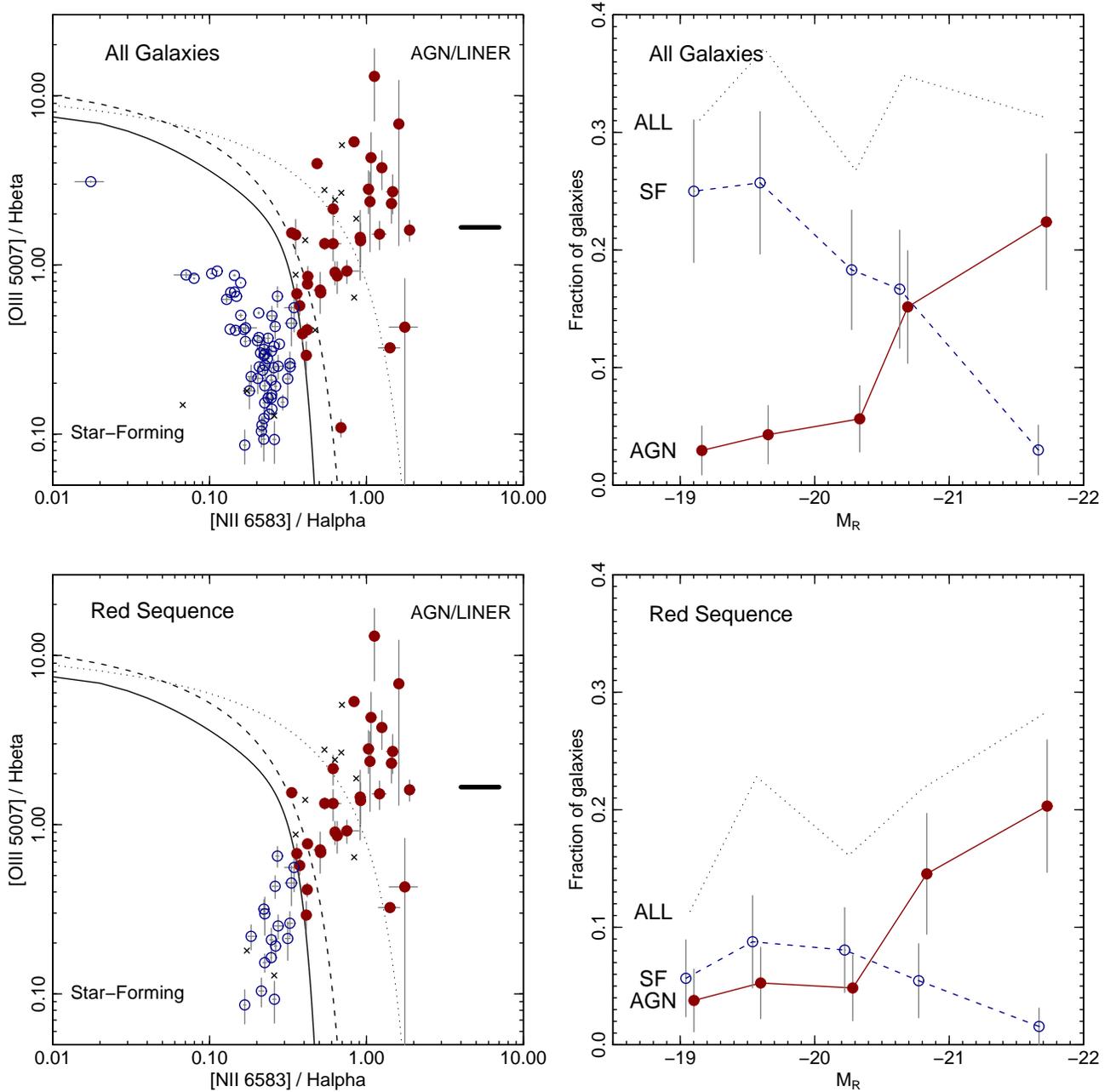}
\caption{Upper left: Emission-line classification diagram, following Baldwin, Phillips
\& Terlevich (1981), showing the 32 per cent of our Shapley sample with measured \ha. 
The discriminatory lines are from from Kewley et al. (2001) (dotted), 
Kauffmann et al. (2003) (dashed) and Stasi\'nska et al. (2006) (solid). Circular symbols
show galaxies from this paper, having sufficient data to classify the likely excitation 
mechanism. Filled circles are galaxies with AGN or LINER emission, according to the Stasi\'nska et al.
(2006) criterion, or to a simple cut at \niib/\ha$>$0.4. Open circles show normal 
star-forming galaxies. Crosses indicate galaxies with detected \ha, but insufficient S/N 
in other lines to permit classification. The thick horizontal bar shows the ratio of \oiiib/\hb\ = $0.6^{-1}$ 
often used in correcting nebular emission in elliptical galaxy studies. 
Upper right panel: fraction of AGN/LINERs, star-forming galaxies, and total fraction with \ha\ 
emission, as a function of total $R$-band magnitude. Lower panels: equivalent diagrams for
the 20 per cent of Shapley red-sequence galaxies with \ha\ emission. 
}
\label{fig:newemiss}
\end{figure*}

Next we consider the origin of the emission lines, i.e whether they 
are due to star formation or to AGN/LINERs. The upper left panel of Figure~\ref{fig:newemiss} 
shows the classical emission-line diagnostic 
plane of Baldwin, Phillips \& Terlevich (1981), hereafter the BPT diagram. 
{(Note that while ratios of line fluxes are usually used in such diagrams, we use ratios of equivalent widths. 
Since we restrict this analysis to line-pairs which are close in wavelength, the ratios differ only at the $\sim0.02$\,dex
level from flux ratios.)}
Normal star-forming galaxies are distinguished from AGN/LINERs by their lower \niib/\ha\ and 
\oiiib/\hb\ ratios. Among the lines proposed to discriminate between the excitation
mechanisms, we plot those from the theoretical modelling of Kewley et al. (2001)
and Stasi\'nska et al. (2006), and the empirical line of Kauffmann et al. (2003), 
derived from SDSS galaxies. The Shapley galaxies populate both the star-forming and the AGN
loci. In classifying the emission source, we adopt the Stasi\'nska et al. criterion, 
which is the `strictest', in the sense of assigning fewest galaxies to the normal star-forming class.
For 23 galaxies with \ha\ emission, we do not detect all of the other three lines
with S/N$>$3 as required for this classification method, but eight of these can be confidently
assigned to the AGN class based on \nii/\ha\ alone. 
Using these criteria we classify 34 galaxies as AGN/LINERs (31 per cent of all the emission-line galaxies), 
and 61 objects (55 per cent) as pure star-forming galaxies. The remaining 15 unclassified galaxies 
have S/N$<$3 in one or more of the lines used, and ambiguous \nii/\ha\ ratio.

The two categories of emission-line galaxy are differently distributed within the luminosity
range spanned by our sample. In the upper right panel of Figure~\ref{fig:newemiss}, we show
the fraction of galaxies (within the NFPS-selected, supercluster-member sample) which have \ha\ emission, 
separated by BPT-classification, as a function of $R$-band luminosity.  Each luminosity bin
represents $\sim$70 galaxies. 
The overall emission-line fraction is approximately constant at $\sim$30 per cent, but the mix between
star-forming galaxies and AGN/LINERs changes dramatically as a function of luminosity. 
Most of the emission population in the highest-luminosity bin are AGN, 
while star-forming objects predominate among the faint emission-line galaxies. The break between
these regimes occurs at $M_R\approx-20.5$, corresponding to luminosity 
$\sim{}10^{10}{\rm L}_\odot$ or a stellar mass $\sim{}10^{10.4}{\rm M}_\odot$. This transition mass
is in excellent agreement with the result obtained by Kauffmann et al. (2003) for a much
larger sample of galaxies from SDSS. 

The above analysis used a sample which includes blue supercluster members. The lower panels of 
Figure~\ref{fig:newemiss} show the same diagrams after imposing an additional selection for only
the red-sequence galaxies ($B-R$ within 0.15 of the fit in Figure~\ref{fig:brcmr}, 291 galaxies).
The overall emission fraction on the red sequence is 20 per cent.
As expected, the star-forming branch is much weaker for the red galaxies. The remaining
galaxies nominally in the star-forming part of the diagram appear to form a continuous sequence with the 
AGN/LINER branch (as also seen in the SDSS sample of Yan et al. 2006). 
The BPT-plane sequence is correlated with luminosity:
the median \oiii/\hb\ for red-sequence galaxies with \ha\ emission declines by a factor of ten (from 2.1 to 0.24), 
from the most to least luminous galaxy bins (cf. the traditional ratio $\sim1.7 = 0.6^{-1}$). 
By contrast the systematic change in \ha/\hb\ is less than a factor of two.
The emission fraction in our red-sequence sample declines to $\sim$10 per cent for fainter galaxies, 
with the AGN/LINERs cutting off around  $M_R\approx-20.5$ as before. However, given 
the systematic trend along the sequence with luminosity, it is 
clear that the AGN threshold luminosity depends quite sensitively on the classification line adopted 
(e.g. using the Kewley et al. line would result in a cut-off at higher luminosity). 
Moreover, the apparently continuous distribution in the BPT plane suggests that imposing a classification
may even not be justified at all, at least for red galaxies. 

To summarize, our analysis reconfirms that the use of a constant factor relative to \oiii\ is not a 
viable means of either correcting for or selecting against nebular emission contamination. We have shown that 
most galaxies with contaminated \hb\ can be readily identified from their \ha\ emission. In the present paper 
we opt not to make an emission correction to the absorption indices, but rather to
use the \ha\ measurments to exclude emitting galaxies from the sample entirely.

\section{The final sample}\label{sec:finalsamp}

\begin{figure}
\includegraphics[angle=270,width=84mm]{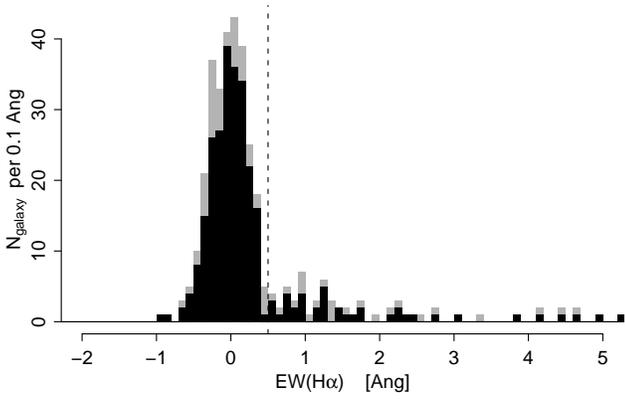}
\caption{Distribution of \ha\ emission equivalent width for the sample galaxies, indicating the imposed limit of 0.5\,\AA. 
Galaxies from our primary NFPS-selected sample are shown as the black histogram, while the grey section indicates the additional
objects from 2MASS. The plot limits are such that most of the emission line galaxies lie outside of the range shown. 
}\label{fig:hadist}
\end{figure}

\begin{figure}
\includegraphics[angle=270,width=84mm]{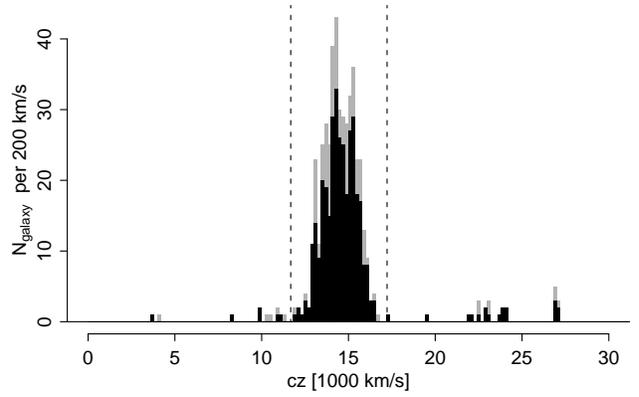}
\caption{Redshift distribution of the observed targets. The black histogram shows the NFPS-selected objects, while the grey 
indicates the contribution from the 2MASS-selected targets. Galaxies with emission lines are included in the histogram.
A total of 57 galaxies lie beyond $z=0.1$ and are not shown. 
The dashed vertical lines indicate the range adopted to define supercluster membership. 
}\label{fig:czdist}
\end{figure}

This section defines explicitly the sample of low-emission supercluster members, 
and tabulates the data for these objects, as analysed in Section~\ref{sec:ixsigs}. 

Recall that after removing galaxies whose entire spectra 
were rejected (stars, superpositions, `fringed' spectra), the sample comprises 540 galaxies, 
with 403 selected from the NFPS imaging and 138 from 2MASS. 
The sample to be analysed in this paper will be the subset of NFPS-selected objects 
having very low levels of nebular emission, and 
having redshifts compatible with membership of the Shapley Supercluster. 
Table~\ref{tab:samplesizes} shows how the various selection criteria affect the sample size. 

The emission selection is based on \ha\ as justified in Section~\ref{sec:emlines}. In defining the low-emission
sample, we use the same cut of 0.5\,\AA\ in EW(\ha), as used in the line-ratio analysis. Figure~\ref{fig:hadist} demonstrates
that this limit is effective in separating the tail of low-emission galaxies from the bulk of apparently emission-free objects. 
Note that the \ha\ cut is equivalent to 0.1--0.2\,\AA\ in \hb, and is thus more restrictive than the 
selection at \hb=0.6\,\AA\ imposed in the NFPS analyses (Nelan et al. 2005; Smith et al. 2006). 

The redshift distribution of sample galaxies is shown in Figure~\ref{fig:czdist}. 
After an iterative $3\sigma$ clipping, the redshift distribution has median 14451\,\kms, with 
a dispersion of 927\,\kms. A total of 448 observed galaxies (342 NFPS plus 106 2MASS) lie within 
the $3\sigma$ range, i.e. 11670--17233\,\kms, which we adopt henceforth as the supercluster redshift limits.
In the foreground, there are two galaxies with redshifts $\sim$4000\,\kms, characteristic of the 
Hydra--Centaurus Supercluster. A few galaxies have velocities $\sim$11000\,\kms, and are excluded by our limits. 
(These are probably associated with a foreground structure including the rich cluster Abell 3571). 
A tail of background galaxies extends to $z\sim0.3$, with a number of groups
apparent, e.g. at $\sim$23000\,\kms, $\sim$27000\,\kms, $\sim$39000\,\kms, $\sim$43000\,\kms, and $\sim$52000\,\kms.
Several of these features have already been noted from wide-field redshift surveys of the region (e.g. Proust et al. 2006). 

The final sample analysed in this paper comprises 232 NFPS-selected galaxies meeting both the emission and the 
redshift criteria, of which 198 have measured velocity dispersions (i.e. are not `unresolved'). 
The 34 unresolved galaxies would have an average of $\sigma=50$\,\kms, if at the median 
of the $M_R-\sigma$ relationship. However, it is likely that they are unresolved precisely because
their $\sigma$ is lower than the average for their luminosity. 

Referring to the solid points in Figure~\ref{fig:brcmr}, we note that although no explicit colour selection has been imposed, the 
redshift and emission cuts result in a sample which is effectively limited to the cluster red sequence. 
A robust fit to the $B-R$ colours yields a slope of $0.043\pm0.005$ per magnitude in $R$, with a (galaxy-by-galaxy) scatter of 0.07\,mag. 
{The culled final sample corresponds to $\sim$35\% of the NFPS $R<18$ input catalogue. Assuming that 
the observed and unobserved galaxies have similar redshift and emission-line distributions, the final sample 
covers $\sim$60\% of all passive supercluster members within the NFPS imaging region. 
}

The median measured velocity dispersion is $\sigma=104$\,\kms, and the 90 per cent range is 39--228\,\kms. The median error in the HgF
index is 0.10\,\AA\ (0.20\,\AA) for velocity dispersions above (below) 100\,\kms. 
{
Since the Thomas et al. (2004) models yield an age-sensitivity of $\partial({\rm HgF})/\partial(\log{\rm Age})\approx-2.6$, these
values correspond to typical formal errors of $\sim$9 per cent ($\sim$20 per cent) in age. 
}

The fully-corrected line-strength data for the final sample are presented in Tables~\ref{tab:basicdat}--\ref{tab:indexdat3}. 

\begin{table*}
\caption{Sample size after application of various selection criteria.}
\label{tab:samplesizes}
\begin{tabular}{llccc}
\hline
\multicolumn{1}{l}{Selection criteria (cumulative)} & Cut applied & \multicolumn{3}{c}{Number of galaxies} \\
& & NFPS-selected & 2MASS-selected & Total \\
\hline
all observed                  &                               & 416 &  149 & 565 \\ 
uncorrupted galaxy spectra    & See Section~\ref{sec:specobs} & 402 &  138 & 540 \\ 
$\lambda$ range for $\sigma$  & $cz\le22000$\,\kms            & 350 &  111 & 461 \\ 
low nebular emission          & EW(\ha)$<$0.5\,\AA            & 232 & \z50 & 282 \\ 
Shapley member                & 11670\,\kms$<cz<$17233\,\kms  & 232 & \z48 & 280 \\   
resolved $\sigma$             & See Section~\ref{sec:sigmas}  & 198 & \z47 & 245 \\   
\hline
\end{tabular}
\end{table*}

\begin{table*}
\caption{Basic data and principal line-strength indices for the sample galaxies. The table includes data for the 
232 galaxies in the NFPS-selected, supercluster-member and emission-free sample as analysed in this paper. 
The galaxy positions can be obtained from their identification numbers. $S/N$ is the signal-to-noise ratio measured
over a rest frame interval of 4500--5500\,\AA. The redshift $cz$ is in the heliocentric frame. 
Columns $R$ and $B$ are total magnitudes from the NFPS photometry. 
Velocity dispersions are given as $\log\sigma$ (with $\sigma$ in \kms), when measured. Galaxies with missing data in this
column are those for which the dispersion was `unresolved'. 
This table includes three non-degenerate absorption line indices Mgb5177, Fe5015 and HgF which can be used to invert stellar 
population models (see Paper II). 
Indices have been corrected to the Lick resolution
and to zero intrinsic velocity broadening. Velocity dispersions and line indices are as observed
through a 2\,arcsec diameter aperture, corresponding to 1.9\,kpc in our adopted cosmology. 
Additional indices are provided in Tables~\ref{tab:indexdat2}~and~\ref{tab:indexdat3}. 
}\label{tab:basicdat}
\begin{tabular}{lcccccccc}
\hline
Galaxy ID & $S/N$ & $cz$ & $R$ & $B$ & $\log\sigma$ & Mgb5177 & Fe5015 & HgF \\
\hline
NFPJ132328.9-314242 & $60$ & $13940$ & $16.127$ & $17.964$ & $2.067\pm0.019$ & $4.107\pm0.124$ & $4.886\pm0.210$ & $-1.384\pm0.144$ \\
NFPJ132330.8-314935 & $93$ & $15061$ & $15.467$ & $17.185$ & $2.014\pm0.012$ & $3.494\pm0.074$ & $4.726\pm0.137$ & $-0.656\pm0.089$ \\
NFPJ132335.5-315201 & $40$ & $14870$ & $16.739$ & $18.577$ & $2.005\pm0.029$ & $3.410\pm0.180$ & $5.576\pm0.307$ & $-1.277\pm0.206$ \\
NFPJ132337.1-315047 & $44$ & $14337$ & $16.163$ & $17.984$ & $2.129\pm0.017$ & $3.812\pm0.152$ & $4.021\pm0.273$ & $-1.144\pm0.170$ \\
NFPJ132345.0-314230 & $54$ & $14999$ & $16.292$ & $18.118$ & $1.843\pm0.039$ & $2.854\pm0.127$ & $4.332\pm0.229$ & $-0.031\pm0.152$ \\
NFPJ132348.3-314953 & $38$ & $15123$ & $17.240$ & $19.061$ & $1.810\pm0.038$ & $3.649\pm0.179$ & $4.617\pm0.316$ & $-0.913\pm0.240$ \\
NFPJ132350.3-313519 & $37$ & $14034$ & $17.330$ & $19.023$ &      --         & $2.611\pm0.196$ & $4.348\pm0.317$ & $\phantom{-}0.331\pm0.235$ \\
NFPJ132355.5-313847 & $34$ & $15248$ & $17.838$ & $19.645$ & $1.861\pm0.046$ & $3.780\pm0.193$ & $4.969\pm0.344$ & $-1.721\pm0.301$ \\
NFPJ132406.9-314449 & $58$ & $13600$ & $16.404$ & $18.260$ & $2.041\pm0.019$ & $3.648\pm0.137$ & $4.426\pm0.220$ & $-1.097\pm0.154$ \\
\hline
\end{tabular}
\\
\scriptsize{The full content of this data table is available in the electronic version of the paper.}
\end{table*}

\begin{table*}
\caption{Supplementary line-strength index data for the sample galaxies.
}\label{tab:indexdat2}
\begin{tabular}{lcccccc}
\hline
Galaxy ID & HdA & HdF & CN1 & CN2 & Ca4227 & G4300 \\
\hline
NFPJ132328.9-314242 & $-1.43\pm0.27$ & $0.56\pm0.19$ & $\phantom{-}0.031\pm0.007$ & $\phantom{-}0.065\pm0.009$ & $1.32\pm0.10$ & $5.60\pm0.22$ \\
NFPJ132330.8-314935 & $-0.77\pm0.15$ & $1.09\pm0.10$ & $\phantom{-}0.016\pm0.004$ & $\phantom{-}0.045\pm0.005$ & $1.06\pm0.06$ & $4.62\pm0.13$ \\
NFPJ132335.5-315201 & $-1.40\pm0.38$ & $0.49\pm0.26$ & $\phantom{-}0.038\pm0.010$ & $\phantom{-}0.077\pm0.012$ & $1.28\pm0.14$ & $5.39\pm0.30$ \\
NFPJ132337.1-315047 & $-1.20\pm0.30$ & $0.35\pm0.21$ & $\phantom{-}0.030\pm0.008$ & $\phantom{-}0.060\pm0.010$ & $0.96\pm0.12$ & $4.94\pm0.26$ \\
NFPJ132345.0-314230 & $\phantom{-}0.68\pm0.24$ & $1.12\pm0.18$ & $-0.034\pm0.007$ & $-0.010\pm0.008$ & $1.07\pm0.09$ & $4.50\pm0.23$ \\
NFPJ132348.3-314953 & $-1.95\pm0.43$ & $0.32\pm0.30$ & $\phantom{-}0.018\pm0.011$ & $\phantom{-}0.056\pm0.014$ & $1.33\pm0.15$ & $5.89\pm0.36$ \\
NFPJ132355.5-313847 & $-0.64\pm0.55$ & $0.88\pm0.39$ & $-0.020\pm0.015$ & $-0.000\pm0.018$ & $1.50\pm0.19$ & $5.36\pm0.44$ \\
NFPJ132406.9-314449 & $-1.12\pm0.28$ & $0.36\pm0.20$ & $\phantom{-}0.016\pm0.008$ & $\phantom{-}0.041\pm0.009$ & $1.24\pm0.11$ & $4.92\pm0.24$ \\
NFPJ132412.7-314658 & $-0.05\pm0.33$ & $1.04\pm0.24$ & $\phantom{-}0.006\pm0.009$ & $\phantom{-}0.037\pm0.011$ & $1.12\pm0.12$ & $4.20\pm0.29$ \\
\hline
\end{tabular}
\\
\scriptsize{The full content of this data table is available in the electronic version of the paper.}
\end{table*}

\begin{table*}
\caption{Supplementary line-strength index data for the sample galaxies.
}\label{tab:indexdat3}
\begin{tabular}{lccccccc}
\hline
Galaxy ID & HgA & Fe4383 & Ca4455 & Fe4531 & Fe4668 & Hbeta & Fe5406 \\
\hline
NFPJ132328.9-314242 & $-5.68\pm0.26$ & $4.67\pm0.28$ & $1.35\pm0.11$ & $3.57\pm0.22$ & $5.91\pm0.30$ & $2.06\pm0.11$ & $1.91\pm0.11$ \\
NFPJ132330.8-314935 & $-4.30\pm0.16$ & $4.58\pm0.17$ & $1.17\pm0.07$ & $3.34\pm0.16$ & $5.55\pm0.19$ & $2.02\pm0.07$ & $1.46\pm0.08$ \\
NFPJ132335.5-315201 & $-5.84\pm0.38$ & $5.31\pm0.40$ & $1.41\pm0.15$ & $3.20\pm0.34$ & $5.33\pm0.44$ & $1.80\pm0.16$ & $1.82\pm0.18$ \\
NFPJ132337.1-315047 & $-5.43\pm0.31$ & $4.45\pm0.35$ & $0.95\pm0.14$ & $2.92\pm0.27$ & $4.60\pm0.38$ & $1.60\pm0.14$ & $1.55\pm0.15$ \\
NFPJ132345.0-314230 & $-3.18\pm0.28$ & $3.89\pm0.31$ & $0.97\pm0.11$ & $2.84\pm0.27$ & $4.29\pm0.35$ & $2.25\pm0.12$ & $1.21\pm0.13$ \\
NFPJ132348.3-314953 & $-5.28\pm0.44$ & $5.18\pm0.44$ & $1.22\pm0.16$ & $2.70\pm0.39$ & $4.72\pm0.48$ & $1.92\pm0.17$ & $1.65\pm0.18$ \\
NFPJ132355.5-313847 & $-5.78\pm0.54$ & $4.61\pm0.52$ & $1.75\pm0.19$ & $3.27\pm0.43$ & $5.14\pm0.52$ & $1.86\pm0.19$ & $1.57\pm0.19$ \\
NFPJ132406.9-314449 & $-4.77\pm0.27$ & $3.86\pm0.32$ & $1.02\pm0.12$ & $2.79\pm0.23$ & $4.37\pm0.31$ & $1.75\pm0.12$ & $1.56\pm0.11$ \\
NFPJ132412.7-314658 & $-4.66\pm0.33$ & $4.84\pm0.36$ & $1.24\pm0.13$ & $3.23\pm0.28$ & $5.66\pm0.40$ & $2.16\pm0.14$ & $1.56\pm0.15$ \\
\hline
\end{tabular}
\\
\scriptsize{The full content of this data table is available in the electronic version of the paper.}
\end{table*}

\section{Results}\label{sec:ixsigs}

In this section, we determine the scaling of each of the Lick line-strength indices as a function of velocity dispersion, 
and use these to infer average scaling relations for age, metallicity and $\alpha$/Fe ratios through comparison to the models of 
Thomas et al. (2003, 2004). The method is essentially that of Nelan et al. (2005). 

\subsection{Index$-\sigma$ relations}

\begin{figure*}
\includegraphics[angle=0,width=180mm]{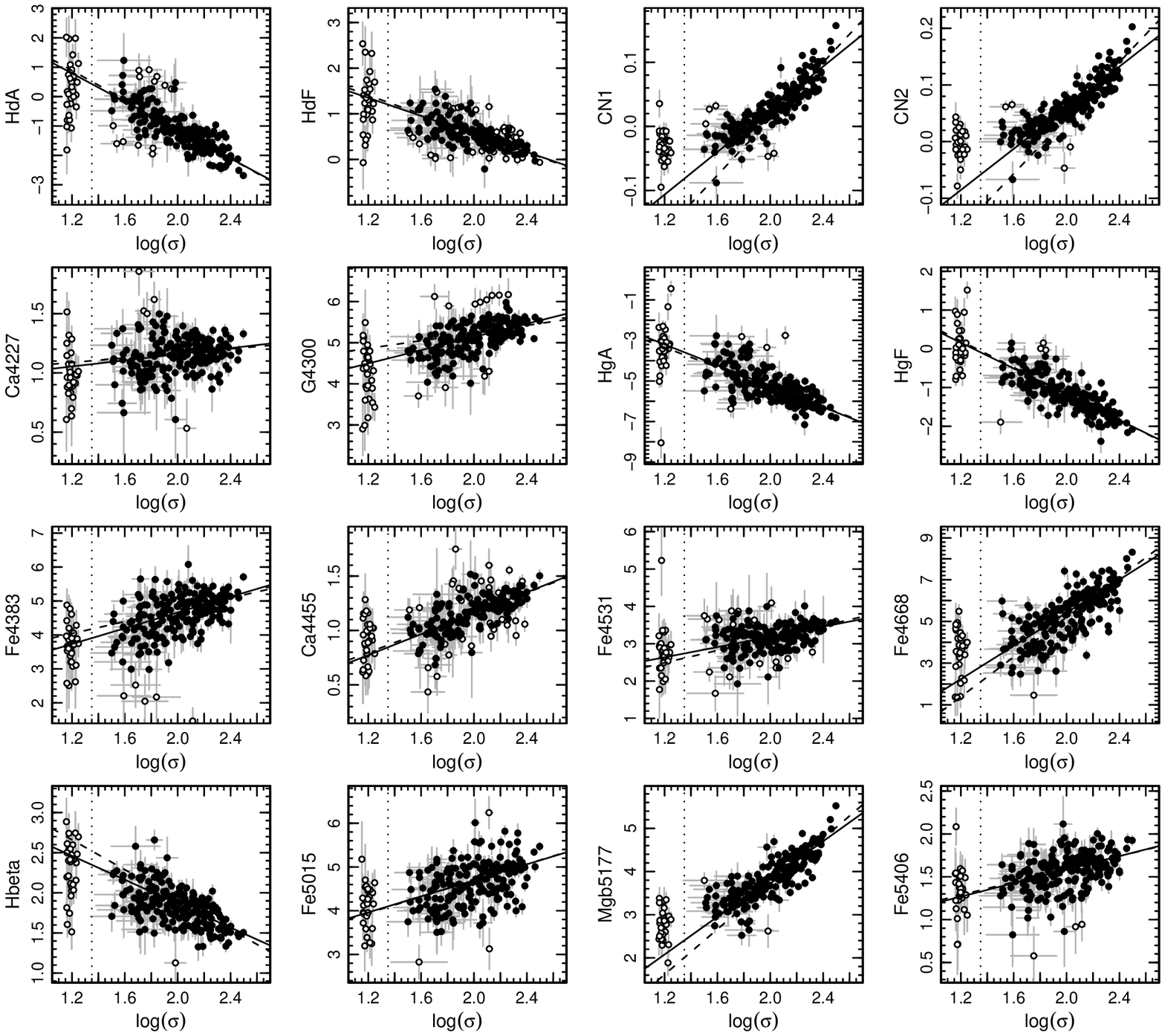}
\caption{Index-$\sigma$ relations. The lines indicate the fits to all of the data (solid) and to the $\sigma>100$\,\kms\ subset (dashed).
The objects with unresolved velocity dispersions (to the left of the dotted line) are not used when fitting the relations. Open symbols
represent these and also other galaxies excluded from the fit by an iterative outlier rejection scheme. 
}\label{fig:ixsigs}
\end{figure*}

\begin{table*}
\caption{Summary of the index$-\sigma$ relations of Figure~\ref{fig:ixsigs}. For each index, the table gives the zero-point and slope
of the fitted relationship, together with robust estimates of the total and intrinsic scatter, $\sigma_{\rm tot}$ and $\sigma_{\rm int}$.
Results are given both for the full sample and for the subset with $\sigma>100$\,\kms.}
\label{tab:ixsigs}
\begin{tabular}{lcccccccc}
\hline
Index & \multicolumn{4}{c}{All galaxies (N=198)} & \multicolumn{4}{c}{$\sigma>100$\,\kms\ (N=104)} \\
      & zero-point & slope & $\sigma_{\rm tot}$ & $\sigma_{\rm int}$ & zero-point & slope &  $\sigma_{\rm tot}$ & $\sigma_{\rm int}$ \\
\hline
HdA        & $+3.691\pm0.276$ & $-2.415\pm0.127$ & $0.382$ & $0.276$ & $+3.888\pm0.605$ & $-2.499\pm0.269$ & $0.317$ & $0.228$ \\
HdF        & $+2.536\pm0.103$ & $-0.989\pm0.048$ & $0.203$ & $0.105$ & $+2.660\pm0.194$ & $-1.044\pm0.086$ & $0.162$ & $0.106$ \\
CN1        & $-0.307\pm0.014$ & $+0.167\pm0.007$ & $0.017$ & $0.013$ & $-0.428\pm0.033$ & $+0.220\pm0.015$ & $0.017$ & $0.015$ \\
CN2        & $-0.303\pm0.016$ & $+0.182\pm0.007$ & $0.021$ & $0.018$ & $-0.450\pm0.037$ & $+0.246\pm0.016$ & $0.020$ & $0.018$ \\
Ca4227     & $+0.899\pm0.083$ & $+0.129\pm0.039$ & $0.135$ & $0.095$ & $+0.973\pm0.183$ & $+0.096\pm0.082$ & $0.114$ & $0.088$ \\
G4300      & $+3.529\pm0.187$ & $+0.805\pm0.086$ & $0.306$ & $0.216$ & $+4.303\pm0.395$ & $+0.466\pm0.174$ & $0.270$ & $0.214$ \\
HgA        & $-0.102\pm0.332$ & $-2.579\pm0.152$ & $0.530$ & $0.425$ & $-0.414\pm0.670$ & $-2.441\pm0.296$ & $0.390$ & $0.323$ \\
HgF        & $+2.158\pm0.180$ & $-1.666\pm0.083$ & $0.272$ & $0.216$ & $+2.249\pm0.376$ & $-1.705\pm0.167$ & $0.202$ & $0.164$ \\
Fe4383     & $+2.340\pm0.283$ & $+1.160\pm0.130$ & $0.485$ & $0.373$ & $+2.991\pm0.588$ & $+0.876\pm0.261$ & $0.334$ & $0.249$ \\
Ca4455     & $+0.196\pm0.060$ & $+0.481\pm0.028$ & $0.119$ & $0.053$ & $+0.247\pm0.133$ & $+0.459\pm0.060$ & $0.095$ & $0.056$ \\
Fe4531     & $+1.816\pm0.164$ & $+0.684\pm0.075$ & $0.280$ & $0.159$ & $+1.504\pm0.346$ & $+0.823\pm0.153$ & $0.208$ & $0.119$ \\
Fe4668     & $-2.536\pm0.535$ & $+3.975\pm0.245$ & $0.800$ & $0.696$ & $-4.330\pm1.187$ & $+4.765\pm0.526$ & $0.724$ & $0.671$ \\
Hbeta      & $+3.354\pm0.111$ & $-0.744\pm0.051$ & $0.214$ & $0.176$ & $+3.778\pm0.231$ & $-0.931\pm0.103$ & $0.185$ & $0.161$ \\
Fe5015     & $+2.865\pm0.279$ & $+0.915\pm0.130$ & $0.433$ & $0.364$ & $+2.828\pm0.708$ & $+0.932\pm0.317$ & $0.351$ & $0.309$ \\
Mgb5177    & $-0.549\pm0.201$ & $+2.188\pm0.093$ & $0.265$ & $0.215$ & $-1.539\pm0.471$ & $+2.627\pm0.210$ & $0.267$ & $0.252$ \\
Fe5406     & $+0.796\pm0.114$ & $+0.394\pm0.053$ & $0.179$ & $0.129$ & $+0.835\pm0.268$ & $+0.377\pm0.120$ & $0.170$ & $0.143$ \\
\hline
\end{tabular}
\end{table*}

For this analysis, we use the sample of low-emission supercluster members,
tabulated in the previous section. 
The `unresolved' galaxies (i.e. those which do not have measured $\sigma$) are, of course, not used in the index$-\sigma$ fits,
leaving a sample of 198 galaxies.
Moreover, on an index-by-index basis, we remove galaxies with errors greater than four times the median (typically only
three or four galaxies). 
The fits were weighted according to the error in the line-strengths (velocity dispersion errors are neglected).
An iterative scheme was used, in which galaxies deviating from the relation by more than four times the intrinsic scatter 
(see below), {\it and} by more than twice their measurement error, were rejected in each step. However, our results are 
not very sensitive to whether this outlier rejection is applied or not.
The total and intrinsic scatter around the relation was derived using outlier-robust methods: 
$\sigma_{\rm tot}$ is the standard deviation of a Gaussian distribution having the same interquartile range as the residuals from 
the index$-\sigma$ relation; $\sigma_{\rm int}$ is the additional Gaussian scatter required, 
in quadrature with the measurement errors, to reproduce the observed interquartile range.
For comparison to earlier work on more luminous galaxy samples, 
the relations were also derived for the 104 galaxies with $\sigma>100$\,\kms.
{ 
We have also fit index$-\sigma$ relations using the traditional VBCs, i.e. a single additive or multiplicative
correction irrespective applied to all galaxies of given $\sigma$. The new VBCs affect the slopes by $\la$3\%, with
the exceptions of: HdA 7\% shallower with new correction, Ca4227 11\% shallower and G4300 8\% steeper. 
}

The index$-\sigma$ relations are presented in Figure~\ref{fig:ixsigs}, and summarized in Table~\ref{tab:ixsigs}. 
Comparison of our $\sigma>100$\,\kms\ slopes with the relations tabulated by Nelan et al. 
reveals that the slopes are typically $\sim$20 per cent flatter in the present work. 
The only indices not consistent with this factor are CN1 and Ca4227.
The difference in slope is in the sense expected from the change in velocity dispersion
scale, but larger than suggested by the $\sigma$ comparison in Section~\ref{sec:sigmas}.
For Ca4227, the resolution corrections applied by Nelan et al. were erroneous, generating
a sharp increase in the index at high $\sigma$ and consequently a much steeper slope. 
Ca4227 is the only case where this effect is non-negligible, since its pass-bands are much narrower 
than any other of the other indices. (Note that Ca4227 was not used in the Nelan et al. stellar population analysis.)
For CN1, where there is apparently some curvature in the relation, 
our slope for the $\sigma>100$\,\kms\ set is slightly steeper than in Nelan et al., 
while our slope for the full sample is flatter. 

\subsection{The stellar population scaling relations}

Consideration of the index$-\sigma$ slopes, and comparison to stellar population models,
provides a means to determine the average scaling of age and metal abundances as a function 
of mass (e.g. Nelan et al. 2005; Clemens et al. 2006). Although using only the slopes forces the scaling
relations to a very simple (and perhaps inaccurate) form, this is an explicitly relative method, and therefore is
insensitive to zero-point offsets in either the data or the models. Further discussion of the merits and
limitations of the method can be found in Nelan et al. and in Smith et al. (2006). 

From the Thomas et al. (2003, 2004) model grids, we determine the responses of each index to the three population 
parameters by fitting a linear relation 
$I = R_{\rm Age} \log {\rm Age} +  R_{{\rm Z}/{\rm H}} [{\rm Z}/{\rm H}] + R_{\alpha/{\rm Fe}} [\alpha/{\rm Fe}]$, 
for models with age 3--15\,Gyr, \afe=0.0-0.5 and \zh=$-0.33, 0.00, +0.35, +0.67$. This is a slight improvement on the 
original method where fiducial values were assumed for two parameters when deriving the index response of the third
parameter. For reference, we provide the new responses in Table~\ref{tab:respo}. 

To recover the scaling relations of the three stellar population parameters, we 
consider the observed slopes for nine of our
index$-\sigma$ relations (HdF, HgF, Hbeta, CN1, Fe4383, Mgb5177, Fe4668, Fe5015 and Fe5406). 
These are chosen to be mainly non-redundant (e.g. only one \hg\ index is used), to avoid indices which
are contaminated by the 5577\,\AA\ sky line in many cases (i.e. Fe5270 and Fe5335), and to avoid
indices with strong dependence on Ca abundance\footnote{Ca is nominally an $\alpha$ element, and
enhanced in the default Thomas et al. (2003, 2004) models, but is observed to track Fe rather than Mg 
(Cenarro et al. 2004, and references therein).}. 
The G4300 index is not used in this analysis since it disagrees strongly with the other indices, and severely 
degrades the fit quality if included. The G4300-$\sigma$ relation is much flatter than expected from the other
indices, perhaps due to the high-metallicity saturation mentioned by Schiavon (2007). 

The scaling relations derived through the comparison of measured slopes with the model-derived responses
are: 
\[
{\rm Age}\,\propto\sigma^{0.52\pm0.06}, \ \  
{\rm Z/H}\,\propto\sigma^{0.34\pm0.04}, \ \  
\alpha/{\rm Fe}\,\propto\sigma^{0.23\pm0.04}.
\]
These scalings can be used to predict the index$-\sigma$ slopes
for a wider range of indices, as given in Table~\ref{tab:respo}, which can be directly compared to the observed slopes. 
The quoted errors are the formal uncertainty in the fit, and do not include modelling uncertainties such as the 
scatter when different indices are used for the fit.
{The latter can be estimated conservatively by repeating the exercise for all non-degenerate ``triplets'' chosen
from the above indices. We impose the condition that each triplet must include one Balmer line, one index from the 
set (Mgb5177, CN1, Fe4668) where $R_{\rm \alpha/Fe}$ has the same sign as $R_{\rm {Z/H}}$ and one from the 
set (Fe4383, Fe5015, Fe5406) where \afe\ and \zh\ responses differ in sign. The scatter in scaling relation exponents, 
computed over all 27 of these triplets is  0.10 for age,  0.07 for \zh\ and 0.06 for \afe, which may be taken as
representative of the systematic errors. As a further test, we can use all six metal lines plus each Balmer line in turn. 
In this case, the age exponents are 0.59 using HdF, 0.44 using HgF and 0.54 using Hbeta, with random errors 0.08 in each case. 
}
{Fitting the scaling relations to the index$-\sigma$ slopes derived using the ``traditional'' VBCs (i.e. a common additive
or multiplicative correction function for all spectral types), the exponents change by less than 0.01 relative to our default case. 
Thus we do not support the claim by Kelson et al. (2006, their Figure 10) that subtle differences in VBC method can dramatically
affect the derived relations.}

Applying the same method to our index$-\sigma$ slopes derived from the high-mass ($\sigma>100$\,\kms) subset, we 
obtain: 
\[
{\rm Age}\,\propto\sigma^{0.64\pm0.12}, \ \ 
{\rm Z/H}\,\propto\sigma^{0.38\pm0.09}, \ \  
\alpha/{\rm Fe}\,\propto\sigma^{0.36\pm0.07}.
\]
The high-$\sigma$ age and \zh\ trends are consistent with the full sample within 1$\sigma$, 
while there is a $\sim$2$\sigma$ change in slope for \afe. The latter suggests a 
flattening of the \afe$-\sigma$ relation towards low masses. 
The difference is driven by the  $\alpha$-sensitive Mgb5177 and CN1 indices. 
In particular a curvature, or change in slope around $\log\sigma=2$,
is visually appreciable in the CN index panels of Figure~\ref{fig:ixsigs}. 
Similar behaviour can be seen in the data of \sanch\ et al. (2006a, figure 6). 

The recovered parameter scaling relations from the above analysis are summarized in the upper section of Table~\ref{tab:scalrels}. 

\begin{table}
\caption{Linear responses for indices derived from model grids. 
$R_{\rm par}$ is the change in the index for a decade change in par, where par is Age, Z/H or $\alpha$/Fe. The
responses are derived from multivariate fits to the Thomas et al. (2003, 2004) models with 
age 3--15\,Gyr, \afe=0.0-0.5, and \zh=$-0.33, 0.00, +0.35, +0.67$. The final column uses the 
scaling relations obtained in Section~\ref{sec:ixsigs} to predict the slope of the index$-\sigma$ slope. 
}\label{tab:respo}
\begin{tabular}{lcccc}
\hline
index & $R_{\rm Age}$ &  $R_{{\rm Z}/{\rm H}}$ &  $R_{{\alpha}/{\rm Fe}}$ & predicted slope\\
\hline
HdA         & $-$3.816 & $-$3.614  & $+$4.707  & $-$2.130 \\
HdF         & $-$1.807 & $-$1.278  & $+$1.954  & $-$0.925 \\
CN1         & $+$0.131 & $+$0.199  & $+$0.059  & $+$0.149 \\
CN2         & $+$0.133 & $+$0.208  & $+$0.067  & $+$0.155 \\
Ca4227      & $+$0.877 & $+$1.193  & $+$0.258  & $+$0.921 \\
G4300       & $+$1.513 & $+$1.876  & $+$1.482  & $+$1.765 \\
HgA         & $-$4.383 & $-$3.974  & $+$3.414  & $-$2.845 \\
HgF         & $-$2.630 & $-$2.116  & $+$0.999  & $-$1.857 \\
Fe4383      & $+$1.976 & $+$3.753  & $-$4.099  & $+$1.361 \\
Ca4455      & $+$0.582 & $+$1.224  & $+$0.062  & $+$0.733 \\
Fe4531      & $+$0.865 & $+$1.559  & $-$0.941  & $+$0.764 \\
Fe4668      & $+$1.884 & $+$7.929  & $+$0.503  & $+$3.791 \\
Hbeta       & $-$1.153 & $-$0.585  & $+$0.285  & $-$0.733 \\
Fe5015      & $+$0.826 & $+$2.713  & $-$1.422  & $+$1.025 \\
Mgb5177     & $+$1.576 & $+$2.971  & $+$2.076  & $+$2.307 \\
Fe5406      & $+$0.448 & $+$1.158  & $-$1.182  & $+$0.355 \\
\hline
\end{tabular}
\end{table}

\section{Discussion}\label{sec:otherwork}

In the lower part of Table~\ref{tab:scalrels}, we tabulate for comparison some results from other recent studies. 
The analyses were based either on the index$-\sigma$ slopes method, or on explicit inversion of model grids. Most of the
studies used the Thomas et al. (2003, 2004) models. Exceptions are Graves et al. (2007) (based on models of Schiavon 2007), 
S\'anchez-Bl\'azquez et al. (2006b) (based on Vazdekis et al. 2007) and Clemens et al. 2006 (based on Annibali et al. 2005). Before
discussing the comparisons we again emphasize that our sample spans a factor of $\sim$6 in velocity dispersion, 
compared to a factor $\sim$2--3 in the other work. 

The most comparable studies, using similar galaxy selection criteria, and similar methods, are those from the NFPS
(Nelan et al. 2005; Smith et al. 2006). The two NFPS analyses were based on slightly different samples and data treatment, 
and the differences between their results probably reflect the level of systematic errors in the method. 
Our results from this paper, for the high-$\sigma$ subset, are within the range spanned by the two NFPS analyses. 
For the full sample, i.e. extending into the low-$\sigma$ regime, the trends are slightly flatter than in NFPS, 
especially for \afe. A possible change in some of the index$-\sigma$ slopes has already been commented on, but 
the change in scaling relations may also reflect the difference in $\sigma$ scales discussed in Section~\ref{sec:sigmas}.
Finally, the inclusion of the low mass galaxies may be compromising a key assumption of the slopes method, 
namely that the grids are linear and parallel over the index range spanned by the data. In Paper II, we will bypass 
this requirement by explicitly inverting the model grids on a galaxy-by-galaxy basis. 

Among other work, results based on samples from SDSS are similar to NFPS and to the present survey, in 
targeting red-sequence or early-type galaxies without detailed visual inspection. The work of
Bernardi et al. (2006) agrees fairly well with our results and with NFPS, although all three parameter
slopes are rather steeper (we have quoted their results using HgF as the age indicator; their
age slope is much steeper when Hbeta is used.) The work of Clemens et al. (2006) agrees in age slope, but
yields much steeper \zh\ and \afe\ scalings, which may result from the different models they use, tracking the C
abundance separately from Mg and Fe. In comparing between NFPS, SDSS and our new sample, 
note that the mass baseline covered by SDSS is a factor of $\sim$3 narrower. 
Most recently, Graves et al. (2007) have obtained a relatively shallow age slope and steeper metallicity slope 
from a colour- and emission-selected SDSS sample. 

The compilation of Thomas et al. (2005) yields scalings close to 
those of NFPS, when a fit is made to their whole sample, rather than excluding young low-mass galaxies in advance as in their paper. 
The work of Caldwell et al. (2003), which was among the first to obtain age estimates in the $\sigma<100$\,\kms\ regime, 
suggests a steeper age slope than recovered here, and a comparable metallicity trend. (Caldwell et al. correct for the $\alpha$-enhancement
effect in deriving the metallicities, but do not report \afe\ measurements for their sample.)
The recent study by S\'anchez-Bl\'azquez et al. (2006b), on the other hand, yields a very flat age$-\sigma$ relation
in the Coma cluster (from $\sim$20 galaxies), although a fairly steep relation appears to hold in their Virgo and field/group galaxies. 
The discrepancy between y S\'anchez-Bl\'azquez et al. and the other results is striking, and remains to be understood.
In particular, results from larger samples in Coma (e.g. Matkovi\'c \& Guzman 2005) should help to resolve this issue. 

On balance, most of the recent line-strength-based studies concur on a significant age trend with $\sigma$, such that
while the stars in the most massive galaxies are almost as old as the universe, the low-$\sigma$ galaxies on average
continued to form stars into the $z<1$ era. For example, taking Age\,$\propto\sigma^{0.5}$, and forcing the 
age at 300\,\kms\ to be 12\,Gyr, our trend requires an average age of 
$\sim$7\,Gyr at $\sigma=100$\,\kms, and $\sim$5\,Gyr at $\sigma=50$\,\kms. 
The corresponding redshift of the last star-formation epoch is $z\approx0.8$ for $\sigma=100$\,\kms, 
and $z\approx0.5$ for $\sigma=50$\,\kms. The implications of these results for high-redshift observations
have been discussed elsewhere (e.g. Smith 2005; Nelan et al. 2005), but can be updated here. 
Specifically, some 50 per cent of the faint galaxies, 
now residing on the red sequence $\sim$2 magnitudes below $L^\star$, were star-forming galaxies at $z\approx0.5$. 
The observed age--mass relation therefore suggests that a deficit of faint red-sequence galaxies should be observed
in high redshift clusters. Early claims for such a deficit (De Lucia et al. 2004; Kodama et al. 2004) have been supported
by more extensive observations (De Lucia et al. 2007; Muzzin et al. 2007; Stott et al. 2007), although contrary results 
have been obtained (Andreon 2006). De Lucia et al. define 
a giant-to-dwarf ratio, $G/D$, on the red sequence of 20 clusters, based on number of red galaxies in luminosity intervals 
corrected for passive evolution. They report an increase of  $G/D$ by a factor of $\sim$3 between $z=0$ and $z=0.75$. We have 
conducted some simple Monte-Carlo experiments using the red sequence luminosity function and the luminosity-$\sigma$ 
correlation, to translate our age-$\sigma$ relation into estimates for the evolution in $G/D$. These tests yield $G/D$ 
increases by a factor of 2--5 at $z=0.75$, for 
Age\,$\propto\sigma^{0.5}$, where the range includes different choices of luminosity function parameters, different normalisations 
of the age--$\sigma$ relation etc. We conclude that the ages measured in present day red-sequence galaxies not only require
a deficit of faint red galaxies in high redshift clusters as observed, but also agree quantitatively in the level of 
of $G/D$ evolution out to $z=0.75$. 

As a closing caveat, we note that although the age--$\sigma$ relation has become a standard diagnostic for 
downsizing, it is usually applied to magnitude-limited samples. At a given $\sigma$, galaxies which are younger than average 
for their mass will be brighter than average, and so preferentially included in such samples, with the result of 
steepening the recovered trends. The impact of the bias depends on the limiting luminosity reached by the sample, and is particularly 
severe for surveys such as SDSS which cover only the exponential cut-off in the $\sigma$ distribution.
This effect was discussed by Kelson et al. (2006), and is investigated in greater detail in Paper II. 

\begin{table*}
\caption{Comparison of parameter-vs-$\sigma$ scaling relations from different index sets in this work, and with recent results from 
large samples ($N_{\rm gal}\ga100$).
To emphasize the different mass ranges covered by different studies, we indicate the median velocity dispersion, $\langle\sigma\rangle$, 
and the 90 per cent range (estimated in some cases). The column headed `span' is the ratio of $\sigma$ spanned by the 90 per cent range.}
\label{tab:scalrels}
\begin{tabular}{lccccccccc}
\hline
Reference                       & source &  Method  & 90\% range $\sigma$ & $\langle\sigma\rangle$ & span & age slope & \zh\ slope & \afe\ slope & note \\
\hline
This work                       & & slopes     &  \z39--228 & 104 & 5.8  & $0.52\pm0.06$ & $0.34\pm0.04$ & $0.23\pm0.04$ & $1$ \\
                                & &            &            &     &      & $\phantom{0.52}\pm0.10$ & $\phantom{0.34}\pm0.07$ & $\phantom{0.23}\pm0.06$ & $2$ \\
\multicolumn{2}{l}{This work ($\sigma>100$\,\kms)}  & slopes     &   104--242 & 152 & 2.3  & $0.64\pm0.12$ & $0.38\pm0.09$ & $0.36\pm0.07$ & $1$ \\
                                & &            &            &     &      & $\phantom{0.64}\pm0.23$ & $\phantom{0.38}\pm0.15$ & $\phantom{0.36}\pm0.13$ & $2$ \\
\hline
Nelan et al. (2005)       & NFPS & slopes     &  \z74--260 & 147 & 3.5  & $0.59\pm0.13$ & $0.53\pm0.08$ & $0.31\pm0.06$ & $3$ \\
Smith et al. (2006)       & NFPS & slopes     &  \z74--260 & 147 & 3.5  & $0.72\pm0.14$ & $0.37\pm0.08$ & $0.35\pm0.07$ & $4$ \\
Graves et al. (2007)      & SDSS & inversion  &  100--250 &  175 & 2.5 & $0.35\pm0.03$ & $0.79\pm0.05$ & $0.36\pm0.04$ & $5$ \\
Bernardi et al. (2006)    & SDSS & inversion  &  120--265  & 188 & 2.2 & 0.81    & 0.58 & 0.39     & $6$ \\
Clemens et al. (2006)     & SDSS & hybrid     &  120--270 & 180 & 2.2 & $\sim$0.8\z\, & 0.76 & 0.74 & $7$ \\  
Thomas et al. (2005)      & Compilation & inversion  &  106--310 & 208 & 2.9 & $0.48\pm0.21$ & $0.35\pm0.09$ & $0.40\pm0.05$ & $8$ \\
Caldwell et al. (2003)    &      & inversion  & 58--270 & 153 & 4.6 & 0.80 & 0.32 & & $9$ \\
\hline
\end{tabular}
{\scriptsize 
Notes: 
$^1$ errors are formal uncertainty from fit to multiple indices; 
$^2$ systematic errors from dispersion among non-degenerate index triplets (see text); \\
$^3$ error includes systematic component; 
$^4$ formal error only; 
$^5$ Hbeta ages for quiescent sample, formal error only;
$^6$ results using HgF (their Table 9); \\
$^7$ metallicity trends reported by Clemens et al., our estimate of age trend from their Figure 10; 
$^8$ our fit (error-weighted) to all data from Thomas et al., formal error only; 
$^9$ metallicity trend as reported, our estimate of age trend from their Table 5. \\
}
\end{table*}

\begin{figure}
\includegraphics[angle=270,width=85mm]{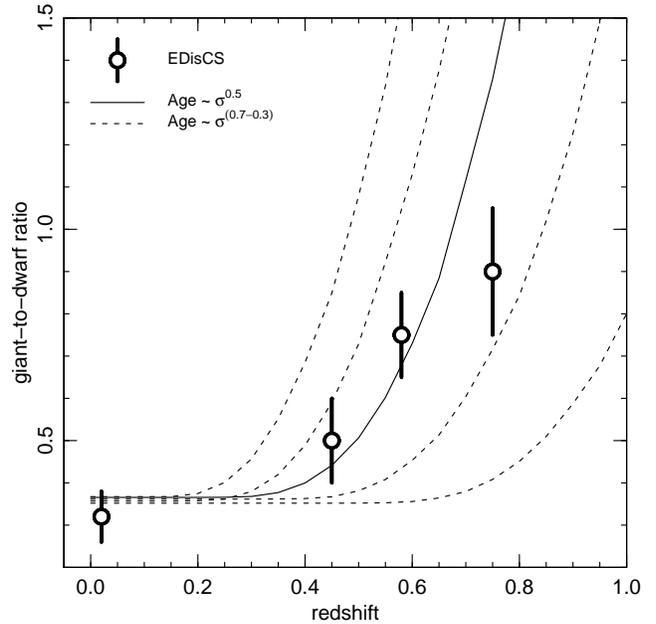}
\caption{Consistency between present day red-sequence ages and the observed 
redshift evolution of the giant-to-dwarf galaxy ratio. The solid line shows
the expected evolution of $G/D$ (defined as in De Lucia et al. 2007), if the 
red-sequence galaxies have Age\,$\propto\sigma^{0.5}$ as found in this paper, 
and if the most massive ellipticals formed their stars at $z\sim3$. The calculation
assumes generic parameters for the luminosity function and the $M_R-\sigma$ 
relation. Dotted lines repeat the calculation for exponents 0.7, 0.6, 0.4, 0.3
(left to right). Points with error bars represent measurements of 
$G/D$ in Coma and in the ESO Distant Cluster Survey (De Lucia et al. 2007). 
}\label{fig:gdr}
\end{figure}

\section{Conclusions}\label{sec:concs}

This paper has presented the first phase of a survey of galaxy properties in the Shapley Supercluster. 
We have described the processing of spectra obtained from $\sim$8\,hr integrations with AAOmega. As well as redshifts
and velocity dispersions, we have measured emission line equivalent widths and absorption line indices. The latter incorporate
an improved treatment of corrections for instrumental resolution and velocity broadening. 

We find a 32 per cent incidence of \ha\ emission in the supercluster sample. For red-sequence galaxies, emission is detected in 20 per cent, 
with line ratios varying systematically from AGN-like at high luminosity to being more consistent with normal star formation at
low luminosity. Defining a low-emission subsample based on the \ha\ measurements, we have presented the 
index$-\sigma$ relations for a range of Lick line-strength indices. We show that the index$-\sigma$ relations and the physical scaling 
relations continue into the low-mass regime with little change in slope in most cases. 
Considering the slopes of these relations,  in the context 
of the Thomas et al. (2003, 2004) stellar population models, we derived 
estimates for the scaling relations of age, total metallicity and $\alpha$-element abundance ratio. 
All three parameters are found to increase significantly 
with $\sigma$, as found also in samples limited to high-mass red-sequence galaxies. There is a slight tendency towards
flatter scalings at low $\sigma$, especially in the case of \afe. Focusing on the age scaling, our 
estimated relation Age\,$\propto\sigma^{0.5}$ is quantitatively consistent with measurements of dwarf-deficient 
red sequences at redshifts $z=0.4-0.8$ e.g. by De Lucia et al. (2007). 

Our study is unique in including a large number of low-luminosity galaxies, with sufficient 
signal-to-noise to obtain meaningful age and abundance estimates on a galaxy-by-galaxy basis. 
This analysis is presented in Paper II (Smith et al. 2007). Future papers will combine the 
AAOmega spectroscopic data with imaging from the Shapley Optical Survey
(Mercurio et al. 2006; Haines et al. 2006) to address stellar population variations as a function of 
morphology and environment. Observational extensions to the spectroscopic survey are also underway. 

\section*{Acknowledgments}

We are grateful to Quentin Parker, Rob Sharp and Scott Croom, for their expertise in preparing and obtaining the 
AAOmega observations, and for their subsequent support. We thank the anonymous referee for a careful and detailed report on this
paper. RJS was supported for this research under the PPARC rolling grant PP/C501568/1 `Extragalactic Astronomy and Cosmology 
at Durham 2005--2010'. 

{}

\label{lastpage}
\end{document}